\documentclass[journal]{IEEEtran}
\usepackage[T1]{fontenc}
\usepackage{cite}
\usepackage{amsmath,amssymb,amsfonts}
\usepackage[ruled,vlined]{algorithm2e}
\usepackage{graphicx}
\usepackage{textcomp}
\usepackage{xcolor}
\usepackage{booktabs}
\usepackage{threeparttable}
\usepackage{multirow}
\usepackage{diagbox}
\usepackage{url}
\usepackage{threeparttable}
\usepackage{array}
\usepackage{textcase}
\usepackage{rotating}
\usepackage{etoolbox}
\makeatletter
\let\MYcaption\@makecaption
\makeatother

\usepackage[font=footnotesize]{subcaption}

\makeatletter
\let\@makecaption\MYcaption
\makeatother

\makeatletter
\def\UrlAlphabet{%
      \do\a\do\b\do\c\do\d\do\e\do\f\do\g\do\h\do\i\do\j%
      \do\k\do\l\do\m\do\n\do\o\do\p\do\q\do\r\do\s\do\t%
      \do\u\do\v\do\w\do\x\do\y\do\z\do\A\do\B\do\C\do\D%
      \do\E\do\F\do\G\do\H\do\I\do\J\do\K\do\L\do\M\do\N%
      \do\O\do\P\do\Q\do\R\do\S\do\T\do\U\do\V\do\W\do\X%
      \do\Y\do\Z}
\def\UrlDigits{\do\1\do\2\do\3\do\4\do\5\do\6\do\7\do\8\do\9\do\0}
\g@addto@macro{\UrlBreaks}{\UrlOrds}
\g@addto@macro{\UrlBreaks}{\UrlAlphabet}
\g@addto@macro{\UrlBreaks}{\UrlDigits}

\patchcmd{\@algocf@start}
  {-1.5em}
  {0pt}
  {}{}

\makeatother

\newcommand{\SNR}[2]{${\text{SNR} #1 \text{#2}\;\text{dB}}$}

\graphicspath{{figures/}}

\def\ng{N_{\rm g}}

\begin{document}
\setlength{\textfloatsep}{5pt}  
\setlength{\floatsep}{5pt}
\renewcommand{\arraystretch}{1.4} 
\ifdefined \GramaCheck
  \newcommand{\CheckRmv}[1]{}
  \newcommand{\figref}[1]{Figure 1}%
  \newcommand{\tabref}[1]{Table 1}%
  \newcommand{\secref}[1]{Section 1}
  \newcommand{\algref}[1]{Algorithm 1}
  \renewcommand{\eqref}[1]{Equation 1}
\else
  \newcommand{\CheckRmv}[1]{#1}
  \newcommand{\figref}[1]{Fig.~\ref{#1}}%
  \newcommand{\tabref}[1]{Table~\ref{#1}}%
  \newcommand{\secref}[1]{Sec.~\ref{#1}}
  \newcommand{\algref}[1]{Algorithm~\ref{#1}}
  \renewcommand{\eqref}[1]{(\ref{#1})}
\fi
\newtheorem{theorem}{Theorem}
\newtheorem{proposition}{Proposition}
\newtheorem{assumption}{Assumption}
\newtheorem{definition}{Definition}
\newtheorem{condition}{Condition}
\newtheorem{property}{Property}
\newtheorem{remark}{Remark}
\newtheorem{lemma}{Lemma}
\newtheorem{corollary}{Corollary}
%
\title{Mini-Batch Gradient-Based MCMC for Decentralized Massive MIMO Detection}
%
%
%
\author{Xingyu~Zhou,~\IEEEmembership{Graduate Student Member,~IEEE,}
        Le~Liang,~\IEEEmembership{Member,~IEEE,}
        Jing~Zhang,~\IEEEmembership{Member,~IEEE,}
		    Chao-Kai~Wen,~\IEEEmembership{Fellow,~IEEE,}
        and~Shi~Jin,~\IEEEmembership{Fellow,~IEEE}
\thanks{This paper was accepted in part for presentation at the International Workshop on Signal Processing Advances in Wireless Communications (SPAWC), Lucca, Italy, in September 2024 \cite{zhou2024decentralized}.}
\thanks{X. Zhou, J. Zhang, and S. Jin are with the National Mobile
Communications Research Laboratory, Southeast University, Nanjing 210096, China
(e-mail: \protect \url{xy_zhou@seu.edu.cn}; jingzhang@seu.edu.cn; jinshi@seu.edu.cn).}
\thanks{L. Liang is with the National Mobile
Communications Research Laboratory, Southeast University, Nanjing 210096, China, and also with Purple Mountain Laboratories, Nanjing 211111, China
(e-mail: lliang@seu.edu.cn).}
\thanks{C.-K. Wen is with the Institute of Communications Engineering,
National Sun Yat-sen University, Kaohsiung 80424, Taiwan
(e-mail: chaokai.wen@mail.nsysu.edu.tw).}
}

%
%

\maketitle
\vspace{-2.0cm}
\begin{abstract}
Massive multiple-input multiple-output (MIMO) technology has significantly enhanced spectral and power efficiency in cellular communications and is expected to further evolve towards extra-large-scale MIMO. 
However, centralized processing for massive MIMO faces practical obstacles, including excessive computational complexity and a substantial volume of baseband data to be exchanged. 
To address these challenges, decentralized baseband processing has emerged as a promising solution. 
This approach involves partitioning the antenna array into clusters with dedicated computing hardware for parallel processing. 
In this paper, we investigate the gradient-based Markov chain Monte Carlo (MCMC) method---an advanced MIMO detection technique known for its near-optimal performance in centralized implementation---within the context of a decentralized baseband processing architecture. 
This decentralized design mitigates the computation burden at a single processing unit by utilizing computational resources in a distributed and parallel manner.
Additionally, we integrate the mini-batch stochastic gradient descent method into the proposed decentralized detector, achieving remarkable performance with high efficiency. 
Simulation results demonstrate substantial performance gains of the proposed method over existing decentralized detectors across various scenarios. 
Moreover, complexity analysis reveals the advantages of the proposed decentralized strategy in terms of computation delay and interconnection bandwidth when compared to conventional centralized detectors.

\end{abstract}


%
\IEEEpeerreviewmaketitle

\section{Introduction}
\IEEEPARstart{M}ASSIVE multiple-input multiple-output (MIMO) technology has emerged as a promising solution to address the increasing demand for high data rates and improved spectral efficiency in wireless communication systems of the fifth generation and beyond \cite{bjornson2017massive}. 
By deploying a large number of antennas at the base station (BS), 
it becomes feasible to achieve a substantial enhancement in network capacity and throughput.
However, the utilization of an extremely large antenna array 
introduces high-dimensional signal processing tasks, particularly challenging in MIMO uplinks, 
where accurate detection of signals from the users is required to fulfill the potential of massive MIMO \cite{yangFiftyYearsMIMO2015,bjornsonMassiveMIMOReality2019}.

Many existing MIMO detection schemes, such as the optimal maximum likelihood (ML) detector and various linear and nonlinear detectors \cite{yangFiftyYearsMIMO2015}, are designed according to the conventional centralized baseband processing architecture. 
However, this architecture faces several challenges when applied to extremely large-scale MIMO systems. 
In this setup, a central unit (CU) assumes responsibility for the entire symbol detection process, imposing stringent demands on the computational capability, baseband interconnection bandwidth, and storage capacity of a single unit \cite{puglielliDesignEnergyCostEfficient2016,liDecentralizedBasebandProcessing2017,rodriguezsanchezDecentralizedMassiveMIMO2020}. 
Notably, existing MIMO detectors inherently involve matrix operations, such as matrix multiplications or inversions, which become impractical on a single computing device as the antenna count increases due to substantial computational complexity.
Additionally, for centralized MIMO detection, the transfer of global channel state information and raw baseband data from all BS antennas to the CU is required. 
This process demands significant interconnection bandwidth, often surpassing data rate limits of existing interconnection standards or chip I/O interfaces \cite{interface2019ecpri,liDecentralizedBasebandProcessing2017,puglielliDesignEnergyCostEfficient2016}.

To address these challenges, decentralized baseband processing (DBP) has been proposed and gained significant attention recently \cite{liDecentralizedBasebandProcessing2017}. 
The concept behind DBP is to partition the large-scale antenna array into a few smaller disjoint antenna clusters, each with local radio frequency (RF) chains and independent baseband processing modules located in distributed units (DUs). 
The DUs handle local baseband data received from RF chains within their respective clusters. 
This DBP approach distributes the demanding signal processing tasks across parallel procedures on multiple computing units, like field-programmable gate arrays \cite{kulkarniHardwareTopologiesDecentralized2021} or graphics processing units \cite{liDecentralizedDataDetection2016}, substantially reducing computation demands at the CU and overall processing latency. 
Moreover, the requirement for raw baseband data transfer between each baseband unit and the corresponding RF circuitry is considerably diminished compared to the centralized architecture, as only a subset of BS antennas needs to transmit their data to the relevant processing module.

Despite the potential advantages of the DBP architecture, research into customized decentralized MIMO detection schemes within this framework remains incomplete. 
Specifically, the exploration of high-performance decentralized detectors is as important as it is currently lacking. 
Pioneering work \cite{liDecentralizedBasebandProcessing2017} decentralized the conjugate gradient algorithm and the alternating direction method of multipliers (ADMM), resulting in iterative consensus-sharing schemes approaching the performance of the centralized linear minimum mean square error (LMMSE) method.
However, the latter is widely known to be suboptimal. 
Similar endeavors were pursued in \cite{kulkarniHardwareTopologiesDecentralized2021} and \cite{liDecentralizedCoordinateDescentData2019} to decentralize the Newton and coordinate descent methods, respectively, aiming to bypass the intricate matrix inversions involved in centralized linear detectors. 
Furthermore, authors of \cite{jeonDecentralizedEqualizationFeedforward2019} developed decentralized feedforward architectures \cite{puglielliDesignEnergyCostEfficient2016,bertilssonScalableArchitectureMassive2016} for classical maximum ratio combining, zero-forcing, and LMMSE detectors, considerably reducing system latency. 
Nonetheless, these linear detectors are markedly suboptimal, and the fully decentralized scheme experiences substantial performance degradation with an increasing number of clusters.

Instead of decentralizing linear detectors, an alternative research direction explores the decentralized design of nonlinear message passing-based detection methods. 
Specifically, authors of \cite{jeonDecentralizedEqualizationFeedforward2019} also applied the large-MIMO approximate message passing (LAMA) algorithm within partially and fully decentralized feedforward architectures. 
However, LAMA's instability arises when channel fading coefficients deviate from the independent and identically distributed (IID) Gaussian distribution. 
In \cite{wangExpectationPropagationDetector2020}, a decentralized detector based on expectation propagation (EP) was proposed, demonstrating performance advantages over various decentralized linear detectors \cite{liDecentralizedBasebandProcessing2017,liDecentralizedDataDetection2016,liDecentralizedCoordinateDescentData2019}. 
Nonetheless, the decentralized EP method necessitates iterative matrix inversions at each DU, posing challenges for achieving computational efficiency with increasing user numbers. 
Other message passing detector variants tailored for the decentralized architecture have been explored \cite{zhangDecentralizedSignalDetection2020,dongEnhancedFullyDecentralized2022,zhangDecentralizedBasebandProcessing2022}. 
Regrettably, these endeavors have yet to strike a desirable balance between performance and complexity.

As evident from the discussion above, existing decentralized MIMO detectors often struggle with suboptimal trade-offs between performance and complexity---particularly in scenarios with high user density or strong correlation channels. 
These limitations underscore the need to explore advanced detection techniques for formulating more competitive decentralized detectors. 
Recently, Markov chain Monte Carlo (MCMC) has demonstrated its potential to overcome the limitations of deterministic Bayesian inference approaches (e.g., EP) \cite{bishopPatternRecognitionMachine,nealMCMCUsingHamiltonian2011}. 
MCMC-based MIMO detectors are favored for their near-optimal performance \cite{farhang-boroujenyMarkovChainMonte2006,hedstromAchievingMAPPerformance2017} and hardware-friendly architecture \cite{larawayImplementationMarkovChain2009}.
Furthermore, the combination of gradient descent (GD) optimization and MCMC offers a framework to efficiently solve the MIMO detection optimum search problem \cite{maSamplingCanBe2019,wellingBayesianLearningStochastic,gowdaMetropolisHastingsRandomWalk2021,zilbersteinAnnealedLangevinDynamics2022}. This scheme employs GD to explore important regions of the discrete search space and utilizes the random walk and acceptance test of MCMC to generate samples conforming to the posterior distribution of transmit symbols \cite{zhouMIMODetectionUsing}. 
Thus, the gradient-based MCMC method holds the potential to achieve near-ML performance, unattainable by LMMSE or message passing-based detectors, while maintaining acceptable complexity. 
Notably, recent work \cite{zhouMIMODetectionUsing} combines MCMC with Nesterov's accelerated gradient (NAG) method \cite{nesterovIntroductoryLecturesConvex2004} to derive a NAG-aided MCMC (NAG-MCMC) detector with near-optimal performance, high efficiency, and excellent scalability. 
However, existing literature has not explored the decentralized design of potent MCMC-based MIMO detectors and their gradient-based variations.

Motivated by the potential benefits of the DBP architecture and the limited exploration of decentralized detectors within this framework, this study seeks to establish a decentralized gradient-based MCMC detector operating within the DBP framework. 
The central goal is to mitigate computational and interconnection bandwidth limitations associated with conventional centralized processing, all while upholding near-optimal detection performance.
The contributions of this paper are summarized as follows:  
\begin{itemize}
  \item 
  \textbf{Decentralized Gradient-Based MCMC Detector for DBP Architecture}: This study pioneers the decentralization of the gradient-based MCMC method for MIMO detection within the DBP architecture, employing both star and daisy-chain topologies. 
  The proposed decentralized detector inherits the remarkable detection performance and scalability inherent in the original gradient-based MCMC algorithm. Decentralization enables distributed and parallel computations, efficiently redistributing the computation burden from the CU to multiple DUs. 
  Consequently, our proposed detector exhibits high computational efficiency in terms of latency.

  \item   \textbf{Mini-Batch-Based Gradient Updating for Efficiency Enhancement}: We introduce a mini-batch-based gradient updating rule for the decentralized gradient-based MCMC detector, resulting in the Mini-NAG-MCMC algorithm. 
  This innovation significantly reduces computation and interconnection demands by selecting only a mini-batch of DUs for gradient computations per iteration. 
  The integration of stochastic gradients within the updating rule synergizes with the MCMC technique, facilitating escape from local minima and achieving near-optimal performance. 
  Additionally, we leverage the asymptotically favorable propagation property of massive MIMO systems, incorporating an approximation into the proposed detector to further reduce overhead.
  
  \item 
  \textbf{Comprehensive Analysis and Performance Validation}: We conduct a comprehensive analysis of the proposed method's computational complexity and interconnection bandwidth, showcasing its superiority over centralized schemes.
  Furthermore, extensive simulations validate the method's performance superiority over existing decentralized detectors across diverse scenarios, including IID {Rayleigh fading} MIMO channels and realistic 3rd Generation Partnership Project (3GPP) channel models. 
  These results affirm the proposed method's effective balance between performance and complexity.

\end{itemize}

\CheckRmv{
	\begin{figure*}[t]
    \setlength{\abovecaptionskip}{-0.15cm}
		\centering
		\includegraphics[width=5.5in]{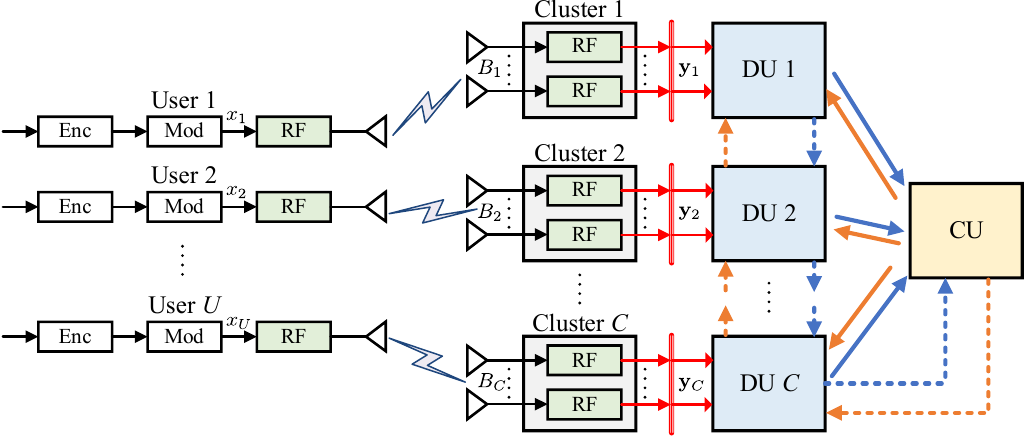}
		\caption{{Block diagram of the uplink massive MIMO system with the DBP architecture.
		The solid lines depict the connections between CU and DUs in the star DBP topology, while the dashed lines depict the counterparts in the daisy-chain DBP topology.}}
		\label{fig:dbp}
	\end{figure*}
}

\textit{Notations:}
Boldface letters denote column vectors or matrices.
$\mathbf{0}$ denotes a zero vector, and $\mathbf{I}$ denotes an identity matrix.
$\mathbf{A}^{T}$, $\mathbf{A}^{\rm H}$, and $\mathbf{A}^{-1}$ denote the transpose, conjugate transpose, and inverse of the matrix $\mathbf{A}$, respectively.
${\rm diag}(\mathbf{A})$ generates a diagonal matrix by assigning zero values to the off-diagonal elements of $\mathbf{A}$.
$\|\cdot\|$ is $l_2$ norm, and $\|\cdot\|_F$ is Frobenius norm.
$\mathbb{E}[\cdot]$ represents the expectation of random variables.
$\mathcal{CN}(\mu,\sigma^2)$ denotes a {circularly symmetric} complex Gaussian distribution with mean $\mu$ and variance $\sigma^2$. $\mathcal{U}(a,b)$ indicates a uniform distribution between $[a,b]$. 
$\mathbb{R}$ and $\mathbb{C}$ are the sets of real and complex numbers, respectively.
The set $[K] = \{1,2,\ldots,K\}$ contains all nonnegative integers up to $K$.

\section{System Model}
In this section, we initially present the uplink MIMO system model and its associated MIMO detection problem. 
Subsequently, we elaborate on the particulars of the DBP architecture for MIMO uplink and discuss the fundamental aspects of designing decentralized MIMO detectors.

\vspace{-0.4cm}
\subsection{Uplink MIMO System Model}
Consider an uplink massive MIMO system with $U$ single-antenna users and a BS equipped with $B\geq U$ receiving antennas, {as shown in \figref{fig:dbp}}.
Each user independently encodes its information bit stream, modulates the coded bits using the quadrature amplitude modulation (QAM) scheme, and  transmits the modulated symbols over the wireless channels to the BS. 
The input-output relationship{\footnote{{We consider a flat-fading channel in the system model; however, the proposed method can be readily generalized to handle frequency-selectivity by using orthogonal frequency division multiplexing (OFDM). In this approach, MIMO detection is independently conducted on each flat-fading subcarrier.}}} 
of this uplink transmission is given by
\CheckRmv{
  \begin{equation}
    \mathbf{y} = \mathbf{Hx} + \mathbf{n},
  \end{equation}
} 
where $\mathbf{y}\in \mathbb{C}^{B \times 1}$ is the received signal at the BS, $\mathbf{H}\in \mathbb{C}^{B \times U}$ is the uplink channel matrix, $\mathbf{x} \in \mathcal{A}^{U \times 1}$ consists of transmit symbols from each user, with $\mathcal{A}$ representing the QAM lattice set having a cardinality of $M$ {and unit symbol energy}, and $\mathbf{n}\in \mathbb{C}^{B \times 1}$ represents the IID {circular complex} Gaussian noise with zero-mean and variance $\sigma^2$ entries. 
The signal-to-noise ratio (SNR) is given by $\text{SNR}=\mathbb{E}[\|\mathbf{Hx}\|^2] / \mathbb{E}[\|\mathbf{n}\|^2]$.
The objective of uplink MIMO detection is to estimate  
the transmit vector $\mathbf{x}$ given the received signal $\mathbf{y}$ and the knowledge of the uplink channel matrix $\mathbf{H}$.  
The optimal ML solution is given by
\CheckRmv{
  \begin{equation}
    {\mathbf{x}}^{\ast}=\underset{\mathbf{x} \in \mathcal{A}^{{U \times 1}}}{\arg \min }\;\|\mathbf{y}-\mathbf{H x}\|^{2}. 
    \label{eq:ml}
  \end{equation}
}
Note that the finite set of $\mathbf{x}$ makes the optimal MIMO detection an integer least-squares problem, known to be non-deterministic polynomial-time hard (NP-hard) \cite{yangFiftyYearsMIMO2015}.  

\subsection{Decentralized Baseband Processing} \label{sec:dbp} 
To deal with the heavy interconnection and computation burdens   
due to the use of extra-large-scale antenna arrays, DBP has been proposed as a promising alternative to traditional centralized processing \cite{liDecentralizedBasebandProcessing2017}. 
Throughout the paper, we assume that the massive MIMO BS possesses the DBP architecture as shown in \figref{fig:dbp}. 
The BS antenna array is partitioned into $C\geq 1$ antenna clusters, where the $c$-th cluster is associated with $B_c$ antennas ($c\in[C]$), and $\sum_{c=1}^C B_c = B$. 
Without loss of generality, we assume that each cluster is of equal size, i.e., 
$B_c = B/C$.  
Each cluster is equipped with local RF chains that are  connected to dedicated hardware referred to as the DU.

We consider two widely accepted DBP topologies, 
namely star and daisy-chain topologies \cite{liDecentralizedBasebandProcessing2017,rodriguezsanchezDecentralizedMassiveMIMO2020,kulkarniHardwareTopologiesDecentralized2021,liDecentralizedGroupwiseExpectation2023,zhaoDecentralizedEqualizationMassive2023}, which are depicted in \figref{fig:dbp}. 
We assume data links among the CU and DUs are bidirectional for both topologies.
The star topology  is featured by all DUs connected to the CU {(solid lines)}.  
Thus, the information exchange between the CU and each DU is in \textit{parallel}.
The daisy-chain topology is featured by DUs connected sequentially 
\cite{rodriguezsanchezDecentralizedMassiveMIMO2020} {(dashed lines)}, with a single connection to the CU at the end. 
Hence, communication is limited solely to adjacent units in the daisy-chain topology  and the information exchange is performed in a \textit{sequential} manner. 
Both these topologies have their distinct strengths in the trade-off among throughput, latency, and interconnection bandwidth \cite{kulkarniHardwareTopologiesDecentralized2021}.
The star topology is well recognized for the low data transfer latency owing to the simultaneous information exchange between the CU and each DU, 
while the daisy-chain topology is known for the constant interconnection bandwidth regardless of the number of DUs, as will be shown in \secref{sec:interconnection}. 

Under the DBP architecture, the uplink channel matrix, received signal, and noise vector are partitioned row-wise as $\mathbf{H} = [\mathbf{H}_1^{T}, \ldots, \mathbf{H}_C^{T}]^T$, $\mathbf{y} = [\mathbf{y}_1^{T}, \ldots, \mathbf{y}_C^{T}]^T$, and $\mathbf{n} = [\mathbf{n}_1^{T}, \ldots, \mathbf{n}_C^{T}]^T$, respectively. 
Hence, the signal model at the $c$-th antenna cluster can be written as
\CheckRmv{
  \begin{equation}
    \mathbf{y}_c = \mathbf{H}_c\mathbf{x} + \mathbf{n}_c,\; c\in [C],
    \label{eq:decentralized_sys}
  \end{equation}
}
where $\mathbf{H}_c\in \mathbb{C}^{B_c \times U}$ is the channel matrix between the $U$ users and the $c$-th antenna cluster. $\mathbf{y}_c \in \mathbb{C}^{B_c\times 1}$ and $\mathbf{n}_c\in \mathbb{C}^{{B}_c \times 1}$ represent the local received signal and noise vector of the $c$-th antenna cluster, respectively.

Each DU $c$ exclusively utilizes 
its local channel matrix $\mathbf{H}_c$ and the received signal $\mathbf{y}_c$
to compute intermediate detection results.\footnote{We assume that the local channel matrix $\mathbf{H}_c$ is perfectly known at each DU since we focus on MIMO detector design in this paper. In realistic scenarios, this knowledge can be obtained by performing local channel estimation.} 
The CU works as a coordinator for periodically aggregating the intermediate information from the DUs and forming the global consensus \cite{liDecentralizedBasebandProcessing2017}, which is then broadcast to the DUs for subsequent updates, and finally producing the estimate.  
In this decentralized scheme, 
the heavy computation burden at the CU can be offloaded to DUs and be performed in a parallel and distributed fashion. 
Hence, the overall computation latency would be significantly reduced. 
Meanwhile, the DUs can process 
the locally stored data to obtain low-dimensional intermediate information instead of directly transmitting full-dimensional data, thus saving interconnection bandwidth.

\section{Mini-Batch Gradient-Based MCMC}
In this section, we first outline the basics of the gradient-based MCMC method from \cite{zhouMIMODetectionUsing}. 
Then, we detail the development of the decentralized gradient-based MCMC detector, Mini-NAG-MCMC, customized for the DBP architecture.

\vspace{-0.1cm}
\subsection{Basics of Gradient-Based MCMC} \label{sec:centralized}
{The MCMC method samples from the target posterior distribution using a Markov chain given by $\mathbf{x}_0, \mathbf{x}_1, \ldots, \mathbf{x}_t, \ldots, \mathbf{x}_S$, where $S$ is the number of sampling iterations. The state transition $\mathbf{x}_{t-1}\to \mathbf{x}_t$ in this chain is regulated by a transition kernel $T(\mathbf{x}|\mathbf{x}_{t-1})$, 
specifically designed to align the chain's stationary distribution with the target distribution.
Hence, the realization of this chain can be utilized to (approximately) generate samples from the target distribution, facilitating statistical inference.
Notably, the integration of MCMC and the GD method has recently emerged as a promising approach for resolving nonconvex optimization challenges \cite{maSamplingCanBe2019}.
The underlying concept of this gradient-based MCMC approach involves leveraging gradients and the Metropolis-Hastings (MH) criterion \cite{hastings1970monte,gowdaMetropolisHastingsRandomWalk2021} to design the transition kernel, expediting the Markov chain's convergence to the target distribution. 
When using gradient-based MCMC for MIMO detection, each state of the Markov chain
corresponds to a sample of $\mathbf{x}$ over the discrete space $\mathcal{A}^{U\times1}$.
The initial sample $\mathbf{x}_0$ is either randomly generated or transferred from another detector. 
The transition from $\mathbf{x}_{t-1}$ to $\mathbf{x}_t$, i.e., the generation of the $t$-th sample $\mathbf{x}_t$, includes the following steps \cite{gowdaMetropolisHastingsRandomWalk2021,zhouMIMODetectionUsing}:}

\subsubsection{GD}  \label{sec:step1}
The constraint $\mathbf{x}\in \mathcal{A}^{U\times 1}$ on the transmit vector can be relaxed to the continuous complex space $\mathbb{C}^{U\times 1}$ to allow the use of gradients.
With this relaxation, the GD method can be performed to provide efficient navigation to important regions of the discrete search space.
Initializing $\mathbf{z}_0 = \mathbf{x}_{t-1}$, where $\mathbf{x}_{t-1}$ is the previous sample, the GD iteration takes the basic form as follows:
\CheckRmv{
  \begin{equation}
    \mathbf{z}_{k} = \mathbf{z}_{k-1} - \tau \nabla f(\mathbf{z}_{k-1}),\; k \in [\ng],
    \label{eq:standard_gd}
  \end{equation}
} 
where $k$ is the index of GD iteration, $N_{\rm g}$ is the total number of GD iterations, $\tau$ is the learning rate, $f(\cdot)$ is the objective function of MIMO detection, which is generally chosen as the residual norm from the estimate, $f(\mathbf{x}) \propto \|\mathbf{y} - \mathbf{Hx}\|^2$,
and $\nabla f(\cdot)$ is the gradient of the objective function.

\subsubsection{Random Walk and QAM Mapping} \label{sec:step2}
After a preset number of GD steps, we generate the MCMC candidate sample $\mathbf{x}^{[\rm cand]}$ as follows:
\CheckRmv{
    \begin{align}
      \mathbf{x}^{[\rm cand]} &= Q(\mathbf{z}_{N_{\rm g}}+ \mathbf{d}), 
      \label{eq:qam_map1}
    \end{align}
}
where $\mathbf{z}_{N_{\rm g}}$ is the output of GD, $\mathbf{d}$ is a random Gaussian vector that represents the perturbation introduced by the random walk, 
and the function $Q(\cdot)$ maps each element of its input to the closest QAM lattice point.
The random walk prevents the algorithm from getting trapped in local minima along the GD direction. 
The QAM mapping discretizes the continuous update to obtain valid candidate samples. 

\subsubsection{Acceptance Test} \label{sec:step3}
The MH criterion \cite{gowdaMetropolisHastingsRandomWalk2021} is applied to accept the candidate sample  
with a probability given by  
\CheckRmv{
  \begin{equation}
    \alpha = \min \left\{1, \frac{\exp (-\|\mathbf{y}-\mathbf{H} \mathbf{x}^{{\rm [cand]}}\|^{2})}{\exp (-\|\mathbf{y}-\mathbf{H} \mathbf{x}_{t-1}\|^{2})}\right\},
    \label{eq:acceptance1}
  \end{equation}
}
{which approximately aligns the Markov chain's stationary distribution with the target posterior distribution.}
This acceptance probability of $\mathbf{x}^{{\rm [cand]}}$ can be achieved by the following rule:
\CheckRmv{
  \begin{equation}
    \mathbf{x}_{t}=\left\{\begin{array}{ll}
      \mathbf{x}^{\rm [cand]}, & \text { if } \alpha \geq \nu \text{ (accept)}, \\
      \mathbf{x}_{t-1}, & \text { otherwise (reject)},
      \end{array}\right.
  \end{equation}
}
where $\nu$ is drawn from the uniform distribution $\mathcal{U}(0,1)$.

Running the gradient-based MCMC sampling for several iterations, a sample list $\mathcal{X}=\{\mathbf{x}_t\}_{t=0}^S$ containing important samples for inference would be identified. 
This sample list can be used to generate the hard decision by choosing the best sample according to 
\CheckRmv{
  \begin{equation}
    \hat{\mathbf{x}} = \underset{\mathbf{x} \in\mathcal{X}}{\arg \min }\;\|\mathbf{y}-\mathbf{H}\mathbf{x}\|^2.
  \end{equation}
}

\CheckRmv{
  \begin{figure*}[t]
    \setlength{\abovecaptionskip}{-0.1cm}
    \centering
    \includegraphics[width=4.5in]{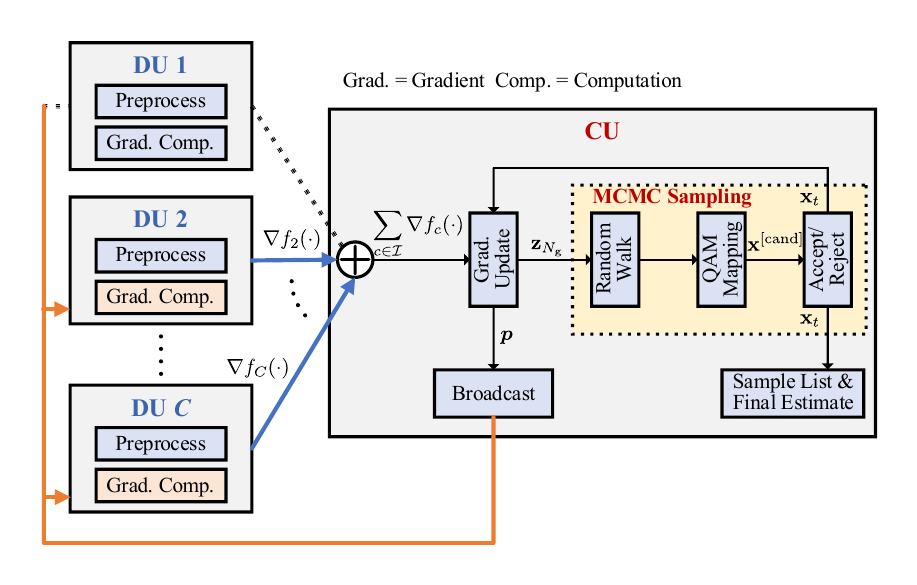}
    \caption{Mini-batch gradient-based MCMC detector implemented using the star DBP topology.  
    The implementaion using the daisy-chain topology  
    is similar and thus omitted.}
    \label{fig:mcmc_dbp}
  \end{figure*}
}

\vspace{-0.6cm}
\subsection{Mini-Batch Gradient-Based MCMC for Decentralized MIMO Detection} \label{sec:mini_grad_mcmc} 
Next, we present the decentralized gradient-based MCMC algorithm tailored for the DBP architecture.
The central concept revolves around distributing the resource-intensive \textit{gradient computations} across multiple DUs, thereby diminishing computation latency and bolstering overall efficiency.

We first provide some useful notations for ease of illustration. 
The local objective function at each DU $c\in [C]$ is defined as 
\CheckRmv{
  \begin{equation}
    f_c(\mathbf{x}) = \frac{1}{2}\|\mathbf{y}_c-\mathbf{H}_c\mathbf{x}\|^2.
    \label{eq:local_obj}
  \end{equation}
}
Therefore, the local gradient is given by
\CheckRmv{
  \begin{equation}
    \nabla f_c(\mathbf{x}) = - \mathbf{H}_c^{\rm H}(\mathbf{y}_c - \mathbf{H}_c\mathbf{x}).
    \label{eq:local_grad}
  \end{equation}
}
The optimal MIMO detection problem in \eqref{eq:ml} is rewritten as  
\CheckRmv{
  \begin{align}
    \mathbf{x}^{\ast}&=\underset{\mathbf{x}\in \mathcal{A}^{U\times 1}}{\arg \min}\; f(\mathbf{x}),
    \label{eq:problem} 
  \end{align}
}
where the global objective function $f(\mathbf{x})$ is given by
\CheckRmv{
  \begin{equation}
    f(\mathbf{x}) \triangleq \frac{1}{2}\|\mathbf{y}-\mathbf{H}\mathbf{x}\|^2 = \sum_{c=1}^{C}f_c(\mathbf{x}).
    \label{eq:global_obj} 
  \end{equation}
}
The decomposition of the global objective function into multiple local objective functions allows the calculation of gradients on each DU as shown in \eqref{eq:local_grad}. 
These local gradients can then be collected at the CU to derive the full gradient $\nabla f(\mathbf{x}) = \sum_{c=1}^{C}\nabla f_c(\mathbf{x})$, which enables the GD-based optimization in \eqref{eq:standard_gd}. 
The distributed computations alleviate  
the computation burden of directly calculating the full gradient at the CU.   

However, the decentralization of the standard full-batch GD in \eqref{eq:standard_gd} does not lead to a reduction in the \textit{overall} computation cost.
Moreover, this decentralized full-batch GD requires information exchange between the CU and \textit{all} DUs in each iteration, resulting in high interconnection costs.
To further mitigate the computation and interconnection requirements associated with the full-batch GD, we turn to
the mini-batch stochastic GD method \cite{bottouOptimizationMethodsLargeScale2018}, which shows promise in efficiently optimizing  objective functions in finite-sum form as \eqref{eq:global_obj}. 
This method serves as an efficient alternative to the full-batch GD in \eqref{eq:standard_gd}, offering the potential for rapid convergence and overhead reduction. 
Specifically, we introduce a mini-batch-based gradient updating rule tailored to the DBP architecture, as detailed below.

In each GD iteration $k$, the estimate is not updated using the full gradient $\nabla f(\cdot)$.  
Instead, a subset of DUs, termed a mini-batch and denoted as $\mathcal{I}$, is randomly selected for the update. 
The mini-batch contains $m=|\mathcal{I}|$ elements.  
Here, $m$ represents the batch size, and we assume that the number of antenna clusters $C$ is divisible by $m$.
The mini-batch-based GD iteration operates as 
\CheckRmv{
  \begin{equation}
    \mathbf{z}_k = \mathbf{z}_{k-1} - \tau \nabla f_{\mathcal{I}}(\mathbf{z}_{k-1}),
    \label{eq:mini_batch_gd}
  \end{equation}
}
where $\nabla f_{\mathcal{I}}(\cdot)$ denotes the estimated gradient based on the mini-batch $\mathcal{I}$. This is expressed as
\CheckRmv{
  \begin{equation}
    \nabla f_{\mathcal{I}}(\mathbf{z}_{k-1}) =  \frac{C}{m}\sum_{c\in \mathcal{I}}\nabla f_c(\mathbf{z}_{k-1}),
    \label{eq:mini_batch_gradient}
  \end{equation} 
}
where the scaling factor $C/m$ ensures the unbiasedness of the mini-batch gradient as an estimator of the full gradient \cite{bottouOptimizationMethodsLargeScale2018}, denoted as, 
\CheckRmv{
  \begin{equation}
    \mathbb{E}[\nabla f_{\mathcal{I}}(\mathbf{x})] = \nabla f(\mathbf{x}),
  \end{equation}
}
with the expectation taken across any possible selections of mini-batch $\mathcal{I}$. 
This guarantees the asymptotic convergence to the minimum of the continuous-relaxed version of problem \eqref{eq:problem} via the mini-batch GD method.

{\begin{remark}
As shown in \eqref{eq:mini_batch_gradient}, only a portion of all DUs are needed for gradient computations and information exchange with the CU per mini-batch GD iteration,
thus providing significant design flexibility for balancing performance and complexity by adjusting the mini-batch size. 
Moreover, simulation results reveal that  
employing mini-batch-based gradients with a sufficiently large mini-batch size (e.g., $m=C/2$) leads to virtually identical  performance compared to using full gradients. 
Hence, the proposed approach enables substantial savings in computational resources and interconnection bandwidth.
\end{remark} 
}

With the proposed mini-batch-based gradient updating rule in mind, we next introduce the {Mini-NAG-MCMC algorithm} for decentralized MIMO detection. 
For ease of presentation, in the remainder of this paper, we consider the star topology as displayed in \figref{fig:mcmc_dbp} unless noted otherwise.
We assume that a total of $S$ samples are generated to perform the final inference for MIMO detection. 
The generation of each sample $\mathbf{x}_t$ is divided into two stages as detailed in the following. 

\subsubsection{Mini-Batch GD Stage} 

In the mini-batch GD stage, we select NAG as the GD component  due to its fastest convergence rate among the first-order gradient-based optimization algorithms \cite{nesterovIntroductoryLecturesConvex2004}.
NAG incorporates momentum in the update rule to expedite and stabilize convergence.
To integrate the NAG method into the DBP architecture, in each GD iteration $k$, 
the incorporation of momentum is performed at the CU to derive the intermediate point $\boldsymbol{p}_k$ as
\CheckRmv{
  \begin{equation}
    \boldsymbol{p}_k = \mathbf{z}_{k-1} + \rho_k \Delta \mathbf{z}_{k-1},
    \label{eq:add_momentum}
  \end{equation}
}
where $\Delta \mathbf{z}_{k-1}=\mathbf{z}_{k-1} - \mathbf{z}_{k-2}$ is the previous update stored at the CU ($\Delta \mathbf{z}_{0}=\mathbf{0}$). 
The momentum factor $\rho_k \in [0,1]$ determines the proportion of the previous update that is added to the current update. 

The intermediate point $\boldsymbol{p}_k$ is then broadcast to DUs belonging to the randomly chosen mini-batch $\mathcal{I}$. 
Upon receiving the broadcast, each DU $c\in \mathcal{I}$ computes the local gradient $\nabla f_c(\boldsymbol{p}_k)$  
in parallel based on \eqref{eq:local_grad}. 
These local gradients are then uploaded and aggregated for performing the descent from $\boldsymbol{p}_k$ and updating the momentum at the CU: 
\CheckRmv{
  \begin{align}
    {\mathbf{z}}_{k}&={\boldsymbol{p}}_{k} -  {\tau\frac{C}{m}}\sum_{c\in \mathcal{I}}\nabla f_c(\boldsymbol{p}_{k}), \label{eq:nag_gd}\\
    \Delta {\mathbf{z}}_{k} &={\mathbf{z}}_{k} - {\mathbf{z}}_{k-1}. 
    \label{eq:update_momentum}%
  \end{align}%
} 
The NAG procedure is iteratively conducted until a preset number of iterations $\ng$ is reached.
The output of NAG, $\mathbf{z}_{\ng}$, is then used as the input for the subsequent MCMC sampling.

One critical parameter that requires careful selection in the NAG is the learning rate $\tau$.
From \cite{nesterovIntroductoryLecturesConvex2004}, a general learning rate choice for an $L$-Lipschitz smooth objective function $f(\cdot)$ is 
\CheckRmv{
  \begin{equation}
    0< \tau \leq \frac{1}{L},
  \end{equation}
} 
where $L$ is the Lipschitz constant, and choosing the upper bound $\tau=\frac{1}{L}$ is beneficial for convergence. 
For the objective function $f(\cdot)$ in \eqref{eq:global_obj}, we have
\CheckRmv{
  \begin{equation}
    L = \lambda_{\max}(\mathbf{G}),
    \label{eq:lipschitz}
  \end{equation}
} 
where $\mathbf{G} = \mathbf{H}^{\rm H}\mathbf{H}$ is the global Gram matrix, and $\lambda_{\max}(\cdot)$ takes the largest eigenvalue of $\mathbf{G}$ \cite{zhouMIMODetectionUsing,odonoghueAdaptiveRestartAccelerated2015}. 
To avoid the computationally expensive eigenvalue decomposition, it is recommended in \cite{zhouMIMODetectionUsing} to replace $\lambda_{\max}(\mathbf{G})$ with its upper bound $\|\mathbf{G}\|_{F}$ and set the learning rate as
\CheckRmv{
  \begin{equation}
    \tau = \frac{1}{\|\mathbf{G}\|_F}.
    \label{eq:lr_fnorm}
  \end{equation}
}

A straightforward way 
to derive $\tau$ in \eqref{eq:lr_fnorm} under the DBP architecture is to accumulate the local Gram matrices from all DUs, forming the global Gram matrix  
at the CU for calculating \eqref{eq:lr_fnorm}. 
Specifically, during the preprocessing period, each DU $c$ computes the local Gram matrix, 
\CheckRmv{
  \begin{equation}
    \mathbf{G}_c=\mathbf{H}_c^{\rm H}\mathbf{H}_c,\; c\in [C],
    \label{eq:local_gram}
  \end{equation}
}
which is then 
collected at the CU to derive the global Gram matrix, $\mathbf{G} = \sum_{c=1}^{C}\mathbf{G}_c$.

Under the so-called favorable propagation in massive MIMO systems \cite{ngoAspectsFavorablePropagation2014}, the computation of $\tau$ can be further simplified.
In such cases, we have
\CheckRmv{
  \begin{equation}
    \frac{1}{B}\mathbf{h}_i^{\rm H}\mathbf{h}_j \to 0,\; i\neq j,\; i, j\in[U],\quad B\to \infty,
  \end{equation}
}
where $\mathbf{h}_i$ denotes the $i$-th column vector of $\mathbf{H}$.
This equation indicates that, for MIMO uplink with a large number of BS antennas, the Gram matrix $\mathbf{G}$ is diagonally dominant and can be approximated by its diagonal counterpart $\mathbf{D}$ given as 
\CheckRmv{
  \begin{equation}
    \mathbf{D} = {\rm diag}(\mathbf{G})=  {\rm diag}\left(\sum_{c=1}^{C}\mathbf{G}_c\right) =\sum_{c=1}^{C}\mathbf{D}_c,
    \label{eq:global_gram_diagonal}
  \end{equation}
}
where $\mathbf{D}_c \in \mathbb{R}^{U\times U}$ is the diagonal counterpart of the local Gram matrix $\mathbf{G}_c$ and can be calculated using the column vector $\mathbf{h}_{u,c}, u\in[U]$ of $\mathbf{H}_c$:
\CheckRmv{
  \begin{equation}
    (\mathbf{D}_{c})_{i j}=\left\{\begin{array}{ll}
      \|\mathbf{h}_{u, c}\|^{2}, & \text { when } i=j=u \\
      0, & \text { otherwise }
      \end{array}\right..
      \label{eq:local_gram_diagonal}
  \end{equation}
}
Finally, the learning rate $\tau$ in \eqref{eq:lr_fnorm} can be approximated as
\CheckRmv{
  \begin{equation}
    \tau \approx \frac{1}{\|\mathbf{D}\|_F}= \frac{1}{\sqrt{\sum_{i=1}^U(\mathbf{D})_{ii}^2}}.
    \label{eq:approx_lr}
  \end{equation}
} 

\begin{remark}
  With the approximation $\mathbf{G}\approx \mathbf{D}$, the complexity on the order of $\mathcal{O}(B_cU^2)$ for calculating \eqref{eq:local_gram} at each DU is substituted by the low complexity of $\mathcal{O}(B_cU)$ for calculating \eqref{eq:local_gram_diagonal}. 
  Meanwhile, only the 
  $U$ real-valued diagonal elements of $\mathbf{D}_c$ are required to be exchanged and accumulated,  
  thereby alleviating the high interconnection bandwidth to convey the entire Gram matrix. 
\end{remark}

\subsubsection{MCMC Sampling Stage} 
After a sufficient GD procedure  
to refine the estimate, 
the candidate sample is generated at the CU by adding random perturbation to ${\mathbf{z}}_{\ng}$  and mapping the perturbed estimate to the QAM lattice using  \eqref{eq:qam_map1}. 
The random Gaussian vector $\mathbf{d}$ in \eqref{eq:qam_map1} is given by
\CheckRmv{
  \begin{equation}
    \mathbf{d} = \gamma \mathbf{M}_{\rm c}\mathbf{w}, 
    \label{eq:random_vector}
  \end{equation}
}
where $\gamma$ is the magnitude of the random perturbation, i.e., step size of the random walk, $\mathbf{M}_{\rm c}\in \mathbb{C}^{U \times U}$ is the Cholesky factor {specifying the covariance of the random perturbation as $\mathbf{M}_{\rm c}\mathbf{M}_{\rm c}^{\rm H}$},  and $\mathbf{w}$ follows $\mathcal{CN}(\mathbf{0}, \mathbf{I})$.  
The step size $\gamma$ is selected to enable sufficient ability to escape local minima while maintaining a reasonable acceptance probability \cite{barbuHamiltonianLangevinMonte2020}. 
Moreover, according to MCMC literature \cite{martinStochasticNewtonMCMC2012,murphy2023probabilistic}, $\mathbf{M}_{\rm c}$ can be set  
to specify a covariance proportional to the inverse of the Hessian matrix of the objective function. 
However, this setting is computationally expensive. 
Experimental observations indicate that using an identity matrix $\mathbf{I}$ for $\mathbf{M}_{\rm c}$ causes almost no performance degradation compared to the former setting, while substantially reducing the computational cost.

Subsequently, the CU performs the MH acceptance test, where the candidate sample $\mathbf{x}^{[\rm cand]}$ is accepted with  
a probability computed according to \eqref{eq:acceptance1}. This probability can be alternatively expressed as
\CheckRmv{
  \begin{align}
    \alpha 
    & =\min \left\{1, \exp \left(2f(\mathbf{x}_{t-1})-2f(\mathbf{x}^{[\mathrm{cand}]})\right)\right\}.
    \label{eq:acceptance2}
  \end{align}
}

Evaluation of the global objective function $f(\cdot)$ 
requires the knowledge of $\mathbf{H}$ and  $\mathbf{y}$, 
which is not directly accessible to the CU. 
Instead, we evaluate the local objective function at each DU and then aggregate them at the CU. 
Specifically, for the acceptance test of the $t$-th sampling iteration, the CU first broadcasts the candidate sample $\mathbf{x}^{\rm [cand]}$ to all the DUs.
Then, each DU independently evaluates the local objective function  in \eqref{eq:local_obj} and uploads the value $f_c(\mathbf{x}^{\rm [cand]})$ to the CU for aggregation to derive 
\CheckRmv{
  \begin{equation}
    f(\mathbf{x}^{\rm [cand]}) = \sum_{c=1}^{C}f_c(\mathbf{x}^{\rm [cand]}).
    \label{eq:global_obj_eval}
  \end{equation}
}
This global objective function value is subsequently used to compute the acceptance probability $\alpha$ in  
\eqref{eq:acceptance2}.  
The value $f(\mathbf{x}^{\rm [cand]})$ is stored or discarded based on the acceptance or rejection of the candidate sample.  
\CheckRmv{
  \begin{equation}
    f(\mathbf{x}_{t})=\left\{\begin{array}{ll}
      f(\mathbf{x}^{\rm [cand]}), & \text { when } \mathbf{x}^{\rm [cand]} \text{ is accepted} \\
      f(\mathbf{x}_{t-1}), & \text { otherwise }
      \end{array}\right..
  \end{equation}
}
Finally, these global objective function values $\{f(\mathbf{x}_t)\}_{t=0}^{S}$  for the drawn samples are used for generating the final estimate. 

\setlength{\algomargin}{0em} 
\SetAlCapHSkip{0em} 
\CheckRmv{
  \begin{algorithm}[!t]
    \SetKwInput{Init}{CU Init.}
    \SetKwInput{Input}{Input}
    \SetKwInput{CU}{CU}
    \SetKwInput{DUone}{Each DU $c\in \mathcal{I}$}
    \SetKwInput{DUtwo}{Each DU $c\in[C]$}
    \SetKwBlock{mcmc}{MCMC sampling stage:}{end}
    \SetKwBlock{nag}{NAG stage:}{end}
    \caption{Mini-NAG-MCMC}
    {
      \small
    \begingroup
    \Input{$\mathbf{H}_c$, $\mathbf{y}_c$, $c\in [C]$, mini-batch size $m$, momentum factor $\rho_k$, random walk step size $\gamma$  and covariance $\mathbf{M}_{\rm c}\mathbf{M}_{\rm c}^{\rm H}$,
    number of  NAG iterations $\ng$, number of sampling iterations $S$. 
    }
    {\textbf{Preprocessing:}\\
    \quad\textbf{Each DU $c\in[C]$:}\\
    \quad\quad Calculate $\mathbf{D}_c$ via \eqref{eq:local_gram_diagonal} and upload its diagonal to the CU. \\
    \quad\textbf{CU:}\\
    \quad\quad Initial sample $\mathbf{x}_0$ and $f(\mathbf{x}_0)$. Calculate the learning rate $\tau$ via \\
    \quad\quad  \eqref{eq:global_gram_diagonal} and \eqref{eq:approx_lr}. \\
    }
    \textbf{Core:}
    \For{$t=1$ \KwTo $S$}
    {
     \setlength\abovedisplayskip{0pt}
     \setlength\belowdisplayskip{0pt}
     \textbf{Mini-batch GD stage:} \\
     \tcp{Perform $N_{\rm g}$ NAG iterations with ${\mathbf{z}}_{0}=\mathbf{x}_{t-1}$.}
     \For{$k=1$ \KwTo $N_{\rm g}$}{
      \setlength\abovedisplayskip{0pt}
      \setlength\belowdisplayskip{0pt}
      \textbf{CU:}\\
      \quad Add momentum to derive $\boldsymbol{p}_k$ via \eqref{eq:add_momentum};\\
      \quad Randomly select a mini-batch $\mathcal{I}$ and broadcast $\boldsymbol{p}_{k}$ to the  \\
      \quad DUs within $\mathcal{I}$. \\
      \textbf{Each DU $c\in \mathcal{I}$:}\\
      \quad Compute local gradients  $\nabla f_c(\boldsymbol{p}_k)$ via \eqref{eq:local_grad} and upload it \\
      \quad to the CU.\\
      \textbf{CU:}\\
      \quad Aggregate $\nabla f_c(\boldsymbol{p}_k)$ to update $\mathbf{z}_k$ and $\Delta\mathbf{z}_k$ via \eqref{eq:nag_gd} and\\ 
      \quad  \eqref{eq:update_momentum}.
     }

     \mcmc{
      \textbf{CU:}\\
     \quad Derive $\mathbf{x}^{[\rm cand]}$ via \eqref{eq:qam_map1} and \eqref{eq:random_vector};\\
     \quad Broadcast $\mathbf{x}^{[\rm cand]}$ to all the DUs. \\
     \textbf{Each DU $c\in[C]$:}\\
     \quad Calculate $f_c(\mathbf{x}^{[\rm cand]})$ via \eqref{eq:local_obj} and upload it to the CU. \\
     \textbf{CU:}\\
    \quad Aggregate $f_c(\mathbf{x}^{[\rm cand]})$ to derive $f(\mathbf{x}^{[\rm cand]})$ via \eqref{eq:global_obj_eval}; \\
    \quad Calculate $\alpha$ via \eqref{eq:acceptance2} 
    and $\nu\sim\mathcal{U}(0,1)$; \\
    \quad If $\alpha \geq \nu$, set $\mathbf{x}_t=\mathbf{x}^{[\rm cand]}$; otherwise, set $\mathbf{x}_t=\mathbf{x}_{t-1}$;\\
    \quad If $\alpha \geq \nu$, set $f(\mathbf{x}_t)=f(\mathbf{x}^{[\rm cand]})$; otherwise, set \\ 
    \quad $f(\mathbf{x}_t)=f(\mathbf{x}_{t-1})$.
    }
    }
    
    {\textbf{Output:} $\hat{\mathbf{x}} = \underset{\mathbf{x} \in\mathcal{X}}{\arg \min }\;f(\mathbf{x})$, where {$\mathcal{X}=\{\mathbf{x}_t\}_{t=0}^{S}$.}} \\
    \endgroup
    }
  \label{alg:mini}
  \end{algorithm}
}

The above two stages are alternatively executed $S$ times 
for the Markov chain to converge to the steady state and derive a sufficient number of samples that follow the posterior distribution, forming the sample list $\mathcal{X}$ for statistical inference at the CU. 
Moreover, parallel samplers can be deployed to reduce sampling latency and correlation. 
In hard decision, the final estimate $\hat{\mathbf{x}}$ can be obtained by selecting the sample from $\mathcal{X}$ that minimizes the objective function $f(\cdot)$. 
Details of the Mini-NAG-MCMC algorithm are summarized in \algref{alg:mini}.
{In the Appendix, we provide the proof of the convergence of the proposed algorithm in a simplified setting.}

\begin{remark}
  It is worth mentioning that the mini-batch gradient-based MCMC method is also applicable to the daisy-chain topology.
  The difference 
  lies in the exchange  
  of information among processing units. 
  Without loss of generality, we assume DU $C$ is  
  within the selected mini-batch in both the star and daisy-chain topologies.
  For the daisy-chain topology, we assume the DUs are connected in the same order as shown in \figref{fig:dbp}. 
  Taking the information upload stage  
  as an instance, DU $C$ in the star topology directly uploads its local gradient to the CU as shown in \figref{fig:mcmc_dbp};  
  By contrast, DU $C$ 
  in the daisy-chain topology 
  needs to accumulate the gradients of its own with those from the former DUs within the mini-batch to derive $\nabla f_{\mathcal{I}}(\cdot)\propto\sum_{c\in \mathcal{I}}\nabla f_c(\cdot)$  
  and then uploads it to the CU.
  This sequential information exchange results in an increased data transfer latency. 
\end{remark}

\begin{remark}
  Note that soft outputs can also be generated based on the sample list, as shown in \cite{farhang-boroujenyMarkovChainMonte2006,hedstromAchievingMAPPerformance2017,zhouMIMODetectionUsing}. 
  The detailed derivations are beyond the scope of this paper and thus omitted. 
\end{remark}

{\begin{remark}
Although the use of multiple computing units in decentralized detection inevitably results in additional energy consumption, the proposed decentralized detector allows for energy-efficient implementation by significantly reducing the computation of each DU.
This enables DUs to operate in a more energy-efficient region with a lower computational density compared to the CU in the centralized scheme.
Moreover, the mini-batch-based gradient updating rule can further increase energy efficiency  
by involving only a subset of DUs in gradient computations per iteration while keeping others inactive.
\end{remark}}

\section{Complexity Analysis}
In this section, we analyze the complexity of the proposed Mini-NAG-MCMC from two  aspects: computational complexity and interconnection bandwidth.

\CheckRmv{ 
\begin{table}[t]
  \captionsetup[table]{skip=0pt}
  \centering
  \begin{threeparttable}
  \caption{{Computational Complexity of Centralized and Decentralized Detection Algorithms}}
  \setlength\tabcolsep{1.5pt} 
  {\begin{tabular}{llcc} 
  \toprule
  \multicolumn{2}{l}{\multirow{2}{*}{\textbf{Algorithms}}} & \multicolumn{2}{c}{\textbf{Computational complexity}}  \\
  \cmidrule{3-4}
            &         & Each DU & CU                               \\
  \hline
  \multirow{6}{*}{\begin{sideways}Decentralized\end{sideways}}&{Mini-NAG-MCMC}               &   $\mathcal{O}\big(B_c U(\ng S + \sqrt{M})\big)$              &      $\mathcal{O}\big(SU(\ng+\sqrt{M})\big)$                              \\
  &{DeADMM $B_c\geq U$}                      &   $\mathcal{O}\big(TU^2+U^3+B_cU^2\big)$   
      &       $\mathcal{O}(TU)$                          \\
  &{DeADMM $B_c<U$}                      &   $\mathcal{O}\big(T(B_c^2+B_cU)+B_c^3\big)$   
      &       $\mathcal{O}(TU)$                          \\               
  &{DeLAMA}         &   $\mathcal{O}\big(T(U^2+\sqrt{M}U)+B_cU^2\big)$               &     $\mathcal{O}(CU)$                             \\
  &{DeEP}                        &   $\mathcal{O}(TU^3+B_cU^2)$              &   $\mathcal{O}\big(TU(\sqrt{M}+C)\big)$                                \\
  &{DeNewton}                        &   $\mathcal{O}(TB_cU+B_cU^2)$              &   $\mathcal{O}(TU^2+U^3)$                                \\
  \midrule
  \multirow{3}{*}{\begin{sideways}Centralized\end{sideways}}&LMMSE                        &  \multicolumn{2}{c}{$\mathcal{O}(U^3+BU^2)$}                    \\
  &$K$-best              &  \multicolumn{2}{c}{$\mathcal{O}(BU^2+K\sqrt{M}U^2)$}                                \\
  &NAG-MCMC                    &  \multicolumn{2}{c}{$\mathcal{O}\big( S(\ng U^2 + \sqrt{M}U+B)  + BU^2 + \sqrt{M}BU \big)$}                                \\
  \bottomrule
  \end{tabular}}
  \label{tab:complexity}
  \begin{tablenotes}[para,flushleft]
  \footnotesize
  {Note: $T$ is the number of iterations of the iterative detection schemes; $K$ is the list size of the $K$-best detector; $\mathbf{M}_{\rm c}=\mathbf{I}$ is used for NAG-MCMC and Mini-NAG-MCMC.}
  \end{tablenotes}
  \end{threeparttable}
\end{table}
}

\subsection{Computational Complexity} \label{sec:computational_complexity}

{\tabref{tab:complexity} presents the computational complexity of Mini-NAG-MCMC compared with various centralized and decentralized detection algorithms.
The decentralized baselines include the decentralized ADMM (DeADMM) \cite{liDecentralizedBasebandProcessing2017}, LAMA (DeLAMA)\footnote{We consider the fully decentralized version of the LAMA detector proposed in \cite{jeonDecentralizedEqualizationFeedforward2019}.} \cite{jeonDecentralizedEqualizationFeedforward2019}, EP (DeEP) \cite{wangExpectationPropagationDetector2020}, and Newton (DeNewton) \cite{kulkarniHardwareTopologiesDecentralized2021} detectors. Centralized LMMSE \cite{wolnianskyVBLASTArchitectureRealizing1998}, $K$-best sphere decoding \cite{guoAlgorithmImplementationKbest2006}, and NAG-MCMC \cite{zhouMIMODetectionUsing} algorithms are also compared. For the considered decentralized schemes, operations in the CU and each DU are counted separately. We only present the count of a single DU to reflect the parallel processing of the DBP architecture. We remark that while this count is lower than the \textit{total} computational cost from all local processing units, it reflects the computation latency of the algorithm.

In the comparison, the complexity of a $U\times U$ matrix inversion is $\mathcal{O}(U^3)$, 
where $\mathcal{O}(\cdot)$ denotes the standard asymptotic notation \cite[Ch. 3.1]{cormen2001introduction} to indicate the order of complexity.
The complexity of Mini-NAG-MCMC is dominated by gradient computations and objective function evaluation, costing $\mathcal{O}(\ng SB_c U)$ and $\mathcal{O}(\sqrt{M}B_c U)$ at each DU, respectively. The other operations in the MCMC sampling stage, such as the random walk, QAM mapping, and acceptance test, incur a modest cost of $\mathcal{O}\big(SU(\ng+\sqrt{M})\big)$ multiplications at the CU. The comparison in \tabref{tab:complexity} shows that Mini-NAG-MCMC is highly efficient due to avoiding high-complexity matrix multiplications and inversions.}

\tabref{tab:complexity} also reveals that the proposed Mini-NAG-MCMC effectively relieves the computation burden at the CU compared to the centralized algorithms by shifting the computationally intensive gradient computations to the DUs.
As a result, the computational complexity at each processing unit 
is independent of the number of BS antennas $B$, which is beneficial for reducing computation latency.
Furthermore, the proposed Mini-NAG-MCMC possesses the advantage that only a portion of the DUs is required to perform computations in each mini-batch GD iteration of the algorithm.  
Thus, the \textit{total} computational cost  
can be saved to a large extent compared to other decentralized schemes.

{Note that the number of iterations 
also has a significant impact on the complexity of the proposed method and other iterative detection schemes in \tabref{tab:complexity}.
However, it is practically difficult to provide an analytical estimate of the number of iterations required to reach a certain performance level. 
Hence, we resort to empirically evaluating the computational complexity of the detectors listed in \tabref{tab:complexity} and meanwhile taking into account the achieved performance.
Detailed results are shown in \secref{sec:simulation}.}

\subsection{Interconnection Bandwidth} \label{sec:interconnection}
{The interconnection bandwidth is measured by the number of bits transferred from and to the CU within a symbol period.
Denote by $\omega$ the bit-width of each real value 
(real/imaginary parts of a complex scalar number) in the implementation. 
Note that the representation of each QAM symbol in the derived samples only requires $\log_2M$ bits, which can be significantly smaller than $\omega$. 
For centralized schemes, the entities required to be uploaded to the CU include the full channel matrix $\mathbf{H}$ and received signal $\mathbf{y}$, resulting in the bandwidth
\CheckRmv{
  \begin{equation}
    C_{\text{centralized}} = 2(B U+B)\omega.
    \label{eq:interconnect_centralized}
  \end{equation}
}

For the star topology-based Mini-NAG-MCMC, the CU receives $UC$ and $2\ng SUm+(S+1)C$ real values during preprocessing and sampling iterations, respectively.
The CU also broadcasts $2\ng SUm$ continuous-valued real numbers and $(S+1)UC$ QAM symbols during the sampling. For the daisy-chain topology-based Mini-NAG-MCMC, the numbers that are required to be transferred are identical to those in the star topology.
However, the CU requires communication with only \textit{one} directly connected DU, 
thereby significantly reducing interconnection costs. 
Hence, the interconnection bandwidth of Mini-NAG-MCMC with star and daisy-chain topologies are respectively given by
\CheckRmv{
  \begin{align}
    C_{\text{Mini}}^{\text{star}} = & 4\ng S U\omega\cdot m +(S+1)U\log_2M\cdot C \nonumber\\ 
    &+ (S+1+U)\omega \cdot C, \label{eq:interconnect_star} \\
    C_{\text{Mini}}^{\text{chain}} = & 4\ng S U\omega+(S+1)U\log_2M + (S+1+U)\omega.\label{eq:interconnect_chain} 
  \end{align}
}
The results of both topologies do not depend on the BS antenna count $B$, a notable advantage over centralized schemes.}

We note that if we consider the interconnection cost on all links in the daisy-chain topology, the total interconnection bandwidth would be comparable to that of the star topology.
However, the daisy-chain architecture can uniformly allocate the bandwidth demand among all links, resulting in a significant reduction in the individual requirements on each link \cite{rodriguezsanchezDecentralizedMassiveMIMO2020}.
Notwithstanding, the price to pay is extended processing latency as a result of the sequential processing architecture.

\section{Simulation Results} \label{sec:simulation}
In this section, we present numerical results to evaluate the proposed Mini-NAG-MCMC in terms of error rate performance, computational complexity, and interconnection cost.

\vspace{-0.3cm}
\subsection{Simulation Setup}
Both IID {Rayleigh fading} and realistic 3GPP 
channel models  are considered in the simulation. 
{The Rayleigh fading MIMO channel has elements that follow the Gaussian distribution $\mathcal{CN}(0,1/B)$.}
{For realistic channels, the 3D channel model outlined in the 3GPP 36.873 technical report \cite{3gpp36873} is utilized. This model has been widely adopted for its accurate representation of real-world channel environments.}
{We consider both uncoded and coded MIMO systems.}
{For uncoded systems}, we evaluate the bit error rate (BER) performance until $5\times 10^7$ bits are transmitted or the number of bit errors exceeds 1000.
{Different independent channel realizations are considered for each symbol transmission.}
{For coded systems, we employ a rate-$3/4$ low-density parity-check (LDPC) code and the OFDM scheme, wherein each user's coded bits are converted to QAM symbols and mapped onto 216 subcarriers for data transmission.
The detector calculates soft outputs using max-log approximation \cite{farhang-boroujenyMarkovChainMonte2006}. 
Belief propagation decoding is then utilized with a maximum number of 10 iterations.
The block error rate (BLER) is evaluated until $10^{5}$ coded blocks are transmitted with independent channel realizations.}

We compare our method with the centralized LMMSE and NAG-MCMC \cite{zhouMIMODetectionUsing} schemes and 
representative decentralized detector baselines as listed in \tabref{tab:complexity}. 
The optimal ML \cite{yangFiftyYearsMIMO2015} and near-optimal $K$-best \cite{guoAlgorithmImplementationKbest2006} detectors are used as performance benchmarks.
For parameter settings of NAG-MCMC and the proposed Mini-NAG-MCMC, the number of NAG iterations is set to $\ng=4$. 
The choice of momentum factor $\rho_k$ follows the recommended setting in \cite{nesterovIntroductoryLecturesConvex2004}:
\CheckRmv{
  \begin{equation}
    \mu_{0}=1, \; \mu_{k}=\frac{1+\sqrt{1+4 \mu_{k-1}^{2}}}{2}, \;\rho_{k}=\frac{\mu_{k-1}-1}{\mu_{k}}, \; k\in[\ng]. \label{eq:sequence}
  \end{equation}
}
The random walk step size is empirically set to $\gamma=0.05$. 
To ensure computational efficiency, we set $\mathbf{M}_{\rm c}=\mathbf{I}$ in \eqref{eq:random_vector}. 
The initial sample $\mathbf{x}_0$ for both NAG-MCMC and Mini-NAG-MCMC is randomly selected from $\mathcal{A}^{U\times 1}$.

\subsection{Convergence Analysis} \label{sec:convergence}

\CheckRmv{
  \begin{figure}[t]
    \captionsetup[subfigure]{aboveskip=-0.5pt,belowskip=-0.5pt}
    \centering
    \begin{subfigure}{3.1in}
      \includegraphics[width=3.1in]{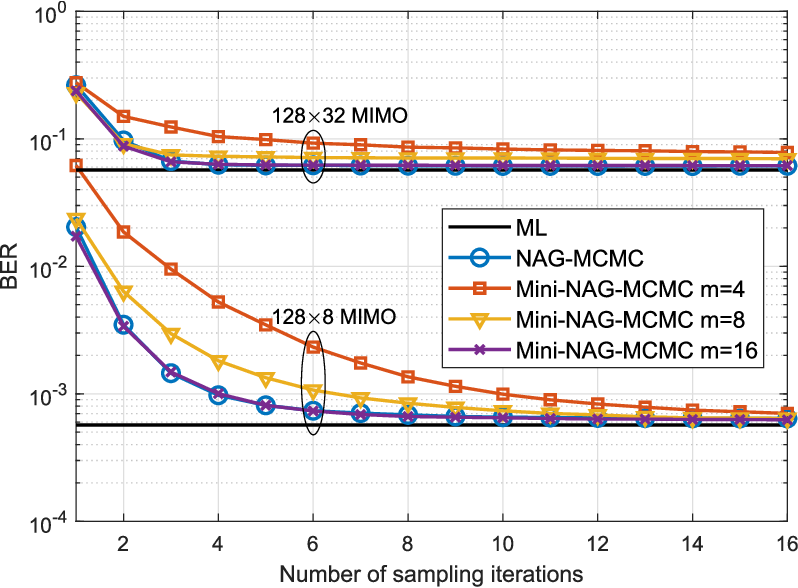}
      \caption{Under different mini-batch sizes $m$ ($C=32$)}
      \label{fig:convergence_m}
    \end{subfigure}
    \begin{subfigure}{3.1in}
      \includegraphics[width=3.1in]{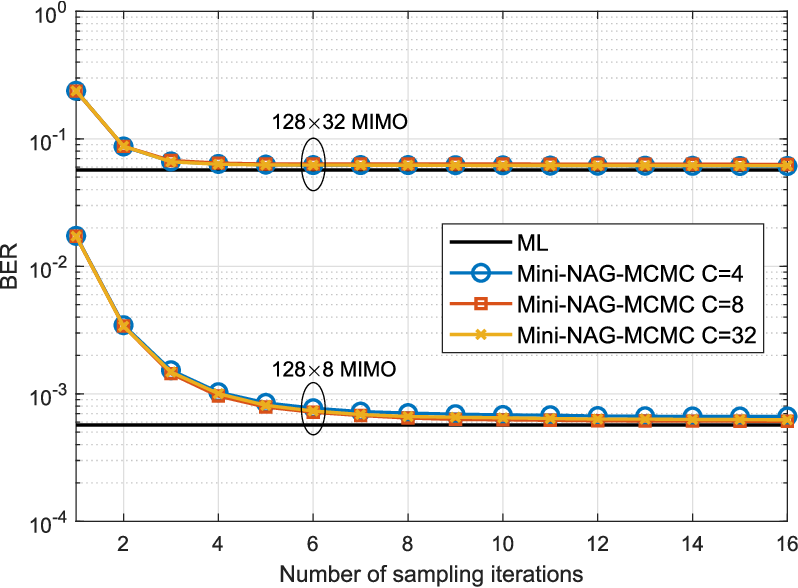}
      \caption{Under different number of clusters $C$ ($m=C/2$)}
      \label{fig:convergence_c}
    \end{subfigure}
    \vspace{-0.1cm}
    \caption{Convergence performance for a MIMO system configuration of $B=128$ and  $U=8$ and $32$ with 16-QAM modulation and \SNR{=}{5} under IID {Rayleigh fading} channels.}
    \label{fig:convergence1}
  \end{figure}
}

\figref{fig:convergence1} presents the convergence performance of Mini-NAG-MCMC in terms of the BER versus the number of sampling iterations.{\footnote{Note that conducting a rigorous theoretical analysis of convergence is challenging for gradient-based MCMC methods in discrete search spaces [22], [26].
Therefore, we have attempted to empirically evaluate the convergence property via numerical simulations, aiming to provide some valuable insights.}} 
The MIMO system is configured with $B=128$ and $U=8$ or $32$. 
The system employs 16-QAM modulation and operates under IID {Rayleigh fading} channels.

Fig.~\ref{fig:convergence1}(a) shows the convergence behavior under different choices of mini-batch sizes.
We also provide the convergence performance of the centralized NAG-MCMC algorithm as a reference. 
The results indicate that Mini-NAG-MCMC converges faster as the mini-batch size $m$ increases in both $U=8$ and $U=32$ cases. 
This result is in accordance with expectation because the variance of the mini-batch gradient approximation in \eqref{eq:mini_batch_gradient} is a decreasing function of $m$ \cite{qianImpactMinibatchSize2020}. 
Meanwhile, the convergence speed is virtually unaffected by the number of users, revealing the robustness of Mini-NAG-MCMC to varying user loads.

Remarkably, the Mini-NAG-MCMC with $m=16$ achieves virtually the same convergence performance as the centralized NAG-MCMC that uses full gradients. 
This phenomenon can be attributed to the sufficient accuracy of the mini-batch approximation when $m$ is sufficiently large (e.g., $m=C/2$ in this simulation) and the synergistic effect of stochastic gradients and MCMC in escaping local minima. 
Consequently, the proposed method achieves equivalent performance to the centralized counterpart that computes full gradients, while offering the benefit of reduced computational cost since only half of all the DUs are utilized for computations per iteration.

Fig.~\ref{fig:convergence1}(b) compares the convergence performance under different numbers of clusters, with the mini-batch size fixed as $m = C/2$.
The figure reveals that the convergence speed shows minimal variation with respect to the  number of clusters, which is a notable advantage of the proposed Mini-NAG-MCMC. 
Hence, the proposed method can accommodate to different levels of decentralization, allowing for trade-offs between the computational complexity at each DU and the number of interconnection interfaces (as well as bandwidth) 
by flexibly adjusting the system parameter $C$ 
according to different requirements.

\subsection{{Uncoded Detection Performance}}

In this subsection, we investigate the proposed method under {Rayleigh fading MIMO channels in uncoded systems}. 
\figref{fig:iid1} provides the BER performance and computational complexity for a massive MIMO system with $B=128$ BS antennas, $U=32$ users, $C=32$ clusters, and 16-QAM modulation. 
The mini-batch size for Mini-NAG-MCMC is $m=16$.
We compare the proposed method with centralized LMMSE \cite{wolnianskyVBLASTArchitectureRealizing1998}, $K$-best \cite{guoAlgorithmImplementationKbest2006}, and NAG-MCMC \cite{zhouMIMODetectionUsing} detectors and two representative decentralized baselines, namely the DeADMM \cite{liDecentralizedBasebandProcessing2017} and DeEP \cite{wangExpectationPropagationDetector2020} detectors.\footnote{For the implementation of the DeADMM scheme, we use the code released at \url{https://github.com/VIP-Group/DBP}.}  
The list size for the $K$-best detector is set to $K=4$, serving as a near-optimal baseline. 
The number of iterations for DeADMM and DeEP is set as $T=50$ and $T=5$, respectively, which is necessitated by the convergence of the algorithm in the considered system setup.

\CheckRmv{
  \begin{figure}[t]
    \captionsetup[subfigure]{aboveskip=-0.5pt,belowskip=-0.5pt}
    \centering
    \begin{subfigure}{3.1in}
      \includegraphics[width=3.1in]{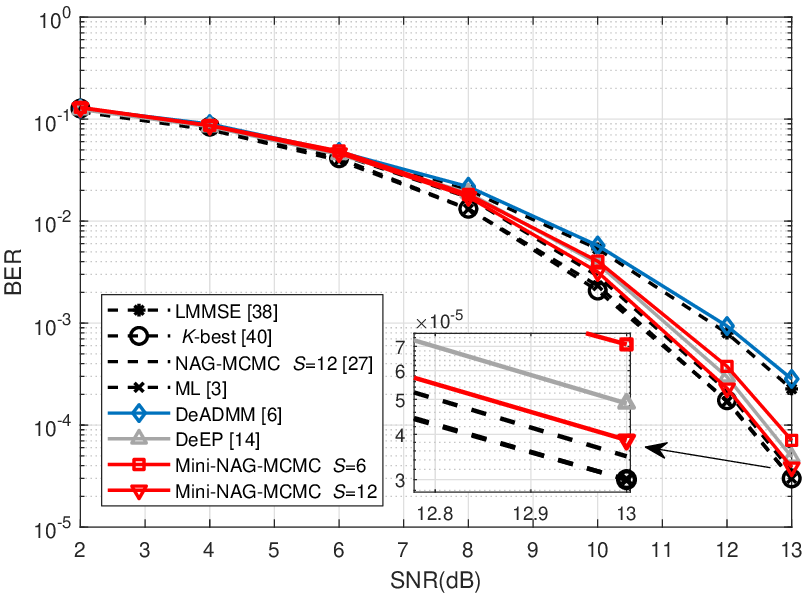}
      \caption{BER performance}
      \label{fig:128x32_ber}
    \end{subfigure}
    \begin{subfigure}{3.1in}
      \includegraphics[width=3.1in]{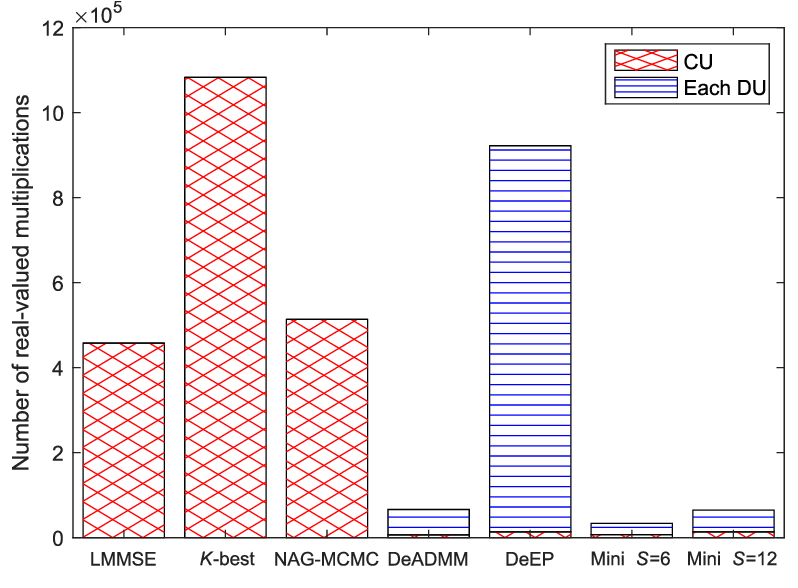}
      \caption{Computational complexity}
      \label{fig:128x32_complexity}
    \end{subfigure}
    \vspace{-0.1cm}
    \caption{BER performance and computational complexity for a MIMO system configuration of $B = 128$, $U = 32$, and $C=32$ with 16-QAM modulation under {Rayleigh fading} channels. 
    The bar chart 
    counts the number of real-valued multiplications at the CU and a single DU for  decentralized schemes.}
    \label{fig:iid1}
  \end{figure}
}

\CheckRmv{
  \begin{figure}[t]
    \captionsetup[subfigure]{aboveskip=-0.5pt,belowskip=-0.5pt}
    \centering
    \begin{subfigure}{3.1in}
      \includegraphics[width=3.1in]{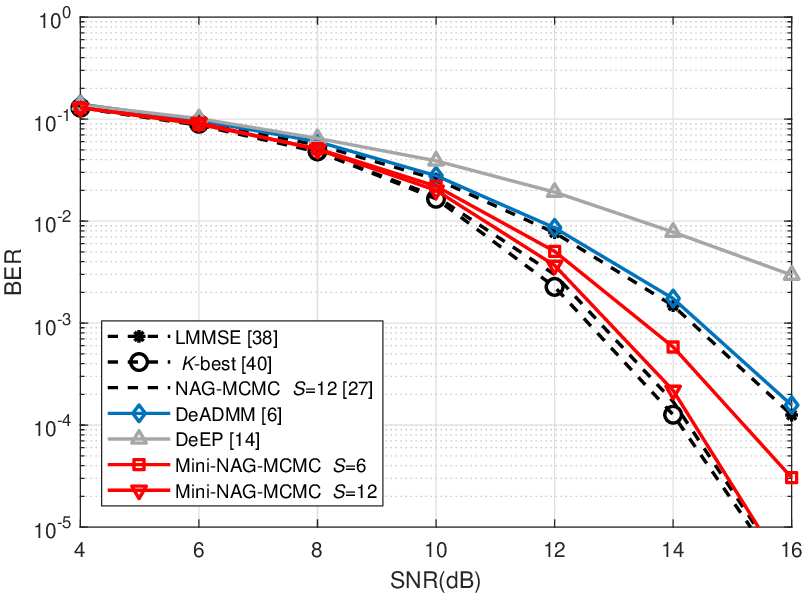}
      \caption{BER performance}
      \label{fig:128x48_ber}
    \end{subfigure}
    \begin{subfigure}{3.1in}
      \includegraphics[width=3.1in]{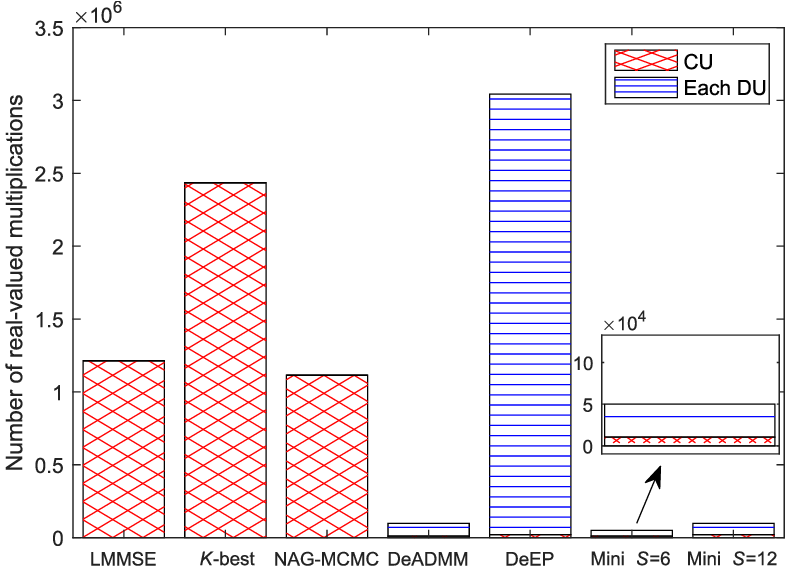}
      \caption{Computational complexity}
      \label{fig:128x48_complexity}
    \end{subfigure}
    \vspace{-0.1cm}
    \caption{BER performance and computational complexity for a MIMO system configuration of $B = 128$, $U = 48$, and $C=32$ with 16-QAM modulation under {Rayleigh fading} channels.}
    \label{fig:iid2}
  \end{figure}
}

Fig.~\ref{fig:iid1}(a) shows that the 
LMMSE and DeADMM detectors have a significant performance gap to the ML detector.
In contrast, Mini-NAG-MCMC  achieves 
a noticeable performance gain over LMMSE and DeADMM when the number of sampling iterations is $S=6$  and further approaches the ML performance with a slight increase in $S$.
Furthermore, Mini-NAG-MCMC  
obtains comparable BER performance to NAG-MCMC when both methods are equipped with the same  $S$.

Fig.~\ref{fig:iid1}(b) illustrates the complexity bar chart of different detectors in terms of the number of real-valued multiplications required per detection. 
The number of multiplications at the CU for Mini-NAG-MCMC is much lower than the counterparts at the CU for the centralized LMMSE, $K$-best, and NAG-MCMC detectors.
We remark that despite the \textit{total} computational cost of Mini-NAG-MCMC being comparable to the centralized LMMSE and NAG-MCMC, the proposed decentralized algorithm offers dual advantages: 
First, the complexity at each DU is notably low, effectively leading to a reduced computation latency.  
Second, the distribution of complexity among many DUs allows the use of cheap computing hardware within each processing unit, in contrast to the centralized algorithms that exert the entire complexity burden on the CU. 
Furthermore, the computational cost of Mini-NAG-MCMC is also similar to that of the DeADMM  and substantially lower than that of the DeEP. 

Considering its near-ML performance and low complexity at each unit, as shown in \figref{fig:iid1}, our proposed Mini-NAG-MCMC demonstrates high promise in achieving an attractive trade-off between performance and complexity.

\figref{fig:iid2} shows the results when the number of users is increased to $U=48$, with the other system settings remaining unchanged compared to \figref{fig:iid1}. 
Fig.~\ref{fig:iid2}(a) indicates that the performance of LMMSE, DeADMM, and DeEP further degrades as compared to the near-optimal $K$-best detector when the user number increases. 
However, Mini-NAG-MCMC demonstrates significant performance improvement over LMMSE when $S=6$ and achieves comparable performance to the $K$-best when $S=12$, consistent with the results presented in Fig.~\ref{fig:iid1}(a).
This finding reflects that the proposed Mini-NAG-MCMC and the associated  learning rate approximation 
are robust to the increased user load. 
Moreover, upon comparing Fig.~\ref{fig:iid2}(b) to Fig.~\ref{fig:iid1}(b), it is observed that the complexity increase of Mini-NAG-MCMC is subtle  in contrast to the LMMSE, $K$-best, and DeEP detectors.
Specifically, the complexity increase is 50\% for Mini-NAG-MCMC, whereas the corresponding values for LMMSE, $K$-best, and DeEP are 165\%, 125\%, and 230\%, respectively,
demonstrating the outstanding scalability of Mini-NAG-MCMC as revealed in \secref{sec:computational_complexity}.

\CheckRmv{
  \begin{figure}[t]
    \captionsetup[subfigure]{aboveskip=-0.5pt,belowskip=-0.5pt}
    \centering
    \begin{subfigure}{3.1in}
      \includegraphics[width=3.1in]{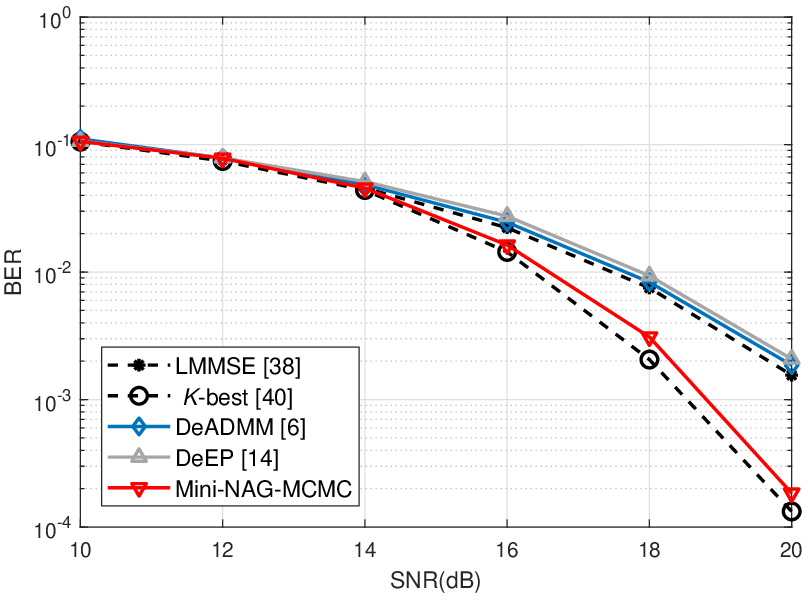}
      \caption{64-QAM}
      \label{fig:128x48_64QAM}
    \end{subfigure}
    \begin{subfigure}{3.1in}
      \includegraphics[width=3.1in]{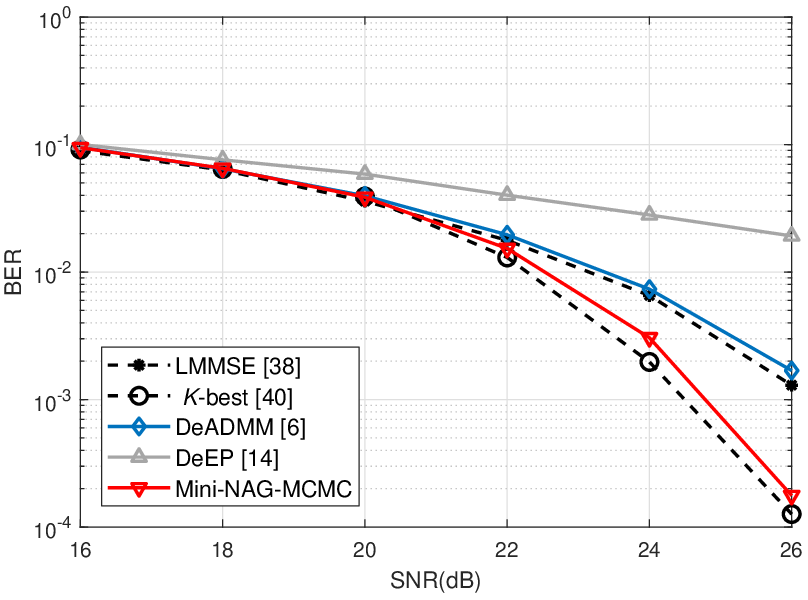}
      \caption{256-QAM}
      \label{fig:128x48_256QAM}
    \end{subfigure}
    \vspace{-0.1cm}
    \caption{BER performance for a MIMO system configuration of $B = 128$, $U = 48$, and $C=32$ with 64-QAM/256-QAM modulation under {Rayleigh fading} channels.}
    \label{fig:iid3}
  \end{figure}
}

\figref{fig:iid3} presents the BER performance under high-order 64-QAM/256-QAM modulation schemes.
The other system configurations are the same as \figref{fig:iid2}.
In this experimental setup, the $K$-best detector is configured with $K=8$ to provide a near-ML baseline.
The DeADMM and DeEP detectors execute for $T=50$ and $T=10$ iterations, respectively.
The proposed Mini-NAG-MCMC is set up with $S=24$ iterations and a mini-batch size of $m=16$.
The figure illustrates  that the proposed method achieves  a performance gain of more than 1.5 dB when compared to the LMMSE detector and the decentralized baselines and approaches the performance of the near-optimal $K$-best.
These results demonstrate that the proposed decentralized detector scales to high-order modulation.

\CheckRmv{
  \begin{figure}[t]
    \captionsetup[subfigure]{aboveskip=-0.5pt,belowskip=-0.5pt}
    \centering
    \begin{subfigure}{3.1in}
      \includegraphics[width=3.1in]{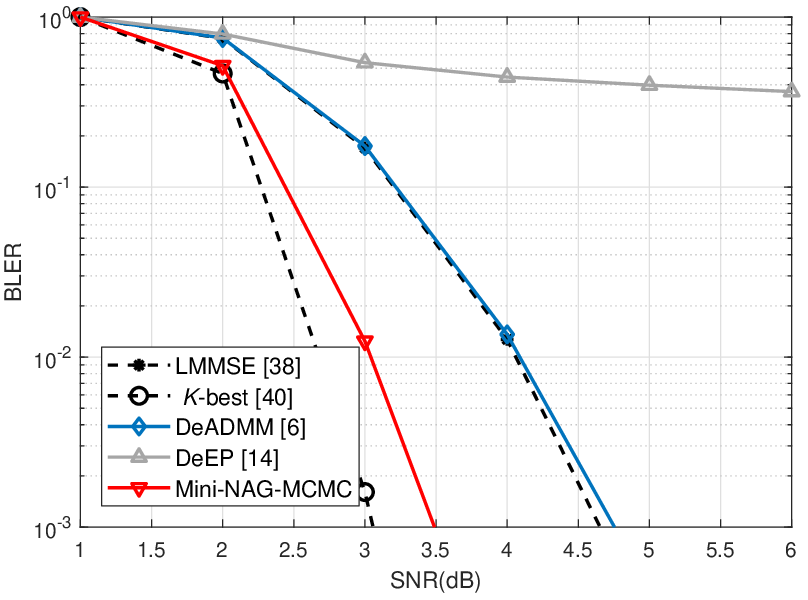}
      \caption{16-QAM}
      \label{fig:bler_3gpp_16qam}
    \end{subfigure}
    \begin{subfigure}{3.1in}
      \includegraphics[width=3.1in]{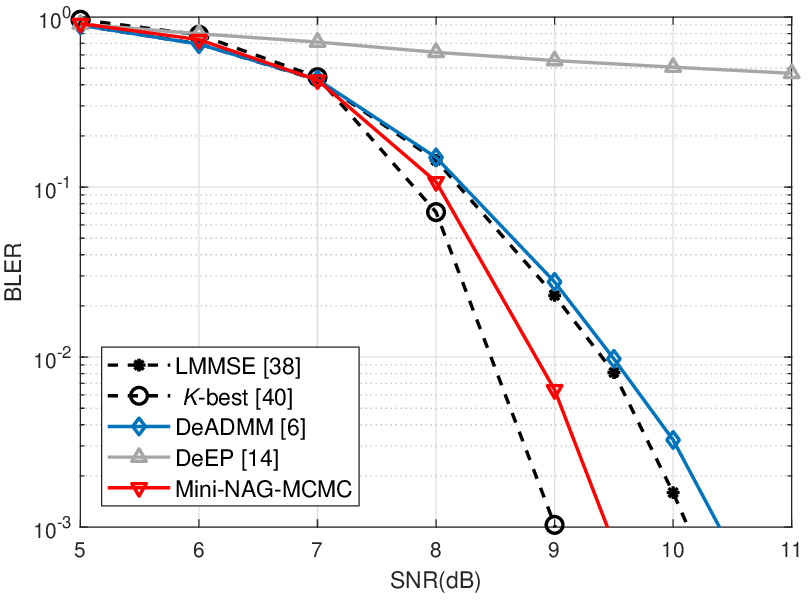}
      \caption{64-QAM}
      \label{fig:bler_3gpp_64qam}
    \end{subfigure}
    \vspace{-0.1cm}
    \caption{{BLER performance for a MIMO system configuration of $B = 128$, $U = 8$, and $C=32$ with 16-QAM/64-QAM modulation and a rate-$3/4$ LDPC code under 3GPP MIMO channels.}}
    \label{fig:bler_3gpp}
  \end{figure}
}

\subsection{{Coded Detection Performance}}
\vspace{-0.1cm}
{We further investigate the proposed method under the more realistic 3GPP 3D MIMO channel model \cite{3gpp36873} with coded transmissions.
We examine the urban macrocell non-line-of-sight scenario.
For the simulation of this scenario, the QuaDRiGa simulator \cite{jaeckel2014quadriga} is employed.}
In the considered scenario, the BS is equipped with $B=128$ antennas, which are uniformly partitioned into $C=32$ clusters, to serve $U=8$ users. 
The BS adopts a uniform planar antenna array with half-wavelength antenna spacing. 
The users are evenly dropped within a $120^{\circ}$ cell sector centered around the BS, spanning a radius ranging from 10 m to 500 m. 
The carrier frequency is set as 2.53 GHz, and the channel spans over 20 MHz bandwidth.
To cope with the complicated channel environments,  the $K$-best detector is set up with $K=8$. The number of iterations for DeADMM and DeEP is selected as $T=50$ and $T=10$, respectively. The number of iterations and mini-batch size for Mini-NAG-MCMC are $S=64$ and $m=16$, respectively.

{As shown in \figref{fig:bler_3gpp}, Mini-NAG-MCMC demonstrates stable performance, exhibiting significant robustness to the realistic channel model.  
Specifically, Mini-NAG-MCMC outperforms LMMSE and other decentralized detectors, achieving an approximate 1 dB gain under both 16-QAM and 64-QAM modulation.
Furthermore, the gap between Mini-NAG-MCMC and the near-optimal performance achieved under centralized schemes by the $K$-best detector is a mere 0.4 dB.  
Considering the distribution of complexity across DUs, achieving near-optimal performance in this context is indeed promising.}

\CheckRmv{
  \begin{figure}[t]
    \setlength{\abovecaptionskip}{-0.2cm}
    \centering
    \includegraphics[width=3.1in]{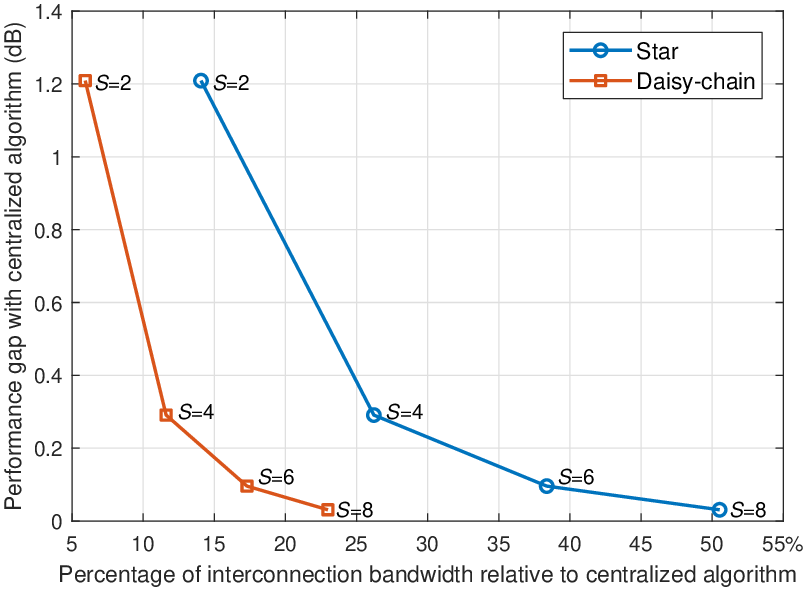}
    \caption{Performance gap versus interconnection bandwidth of Mini-NAG-MCMC ($m=2$ and $S\in [2,4,6,8]$) for a MIMO system configuration of $B = 256$, $U = 8$, and $C=8$ with 16-QAM modulation under {Rayleigh fading} channels.}
    \label{fig:interconnect1} 
  \end{figure}
}

\CheckRmv{
  \begin{figure}[t]
    \setlength{\abovecaptionskip}{-0.2cm}
    \centering
    \includegraphics[width=3.1in]{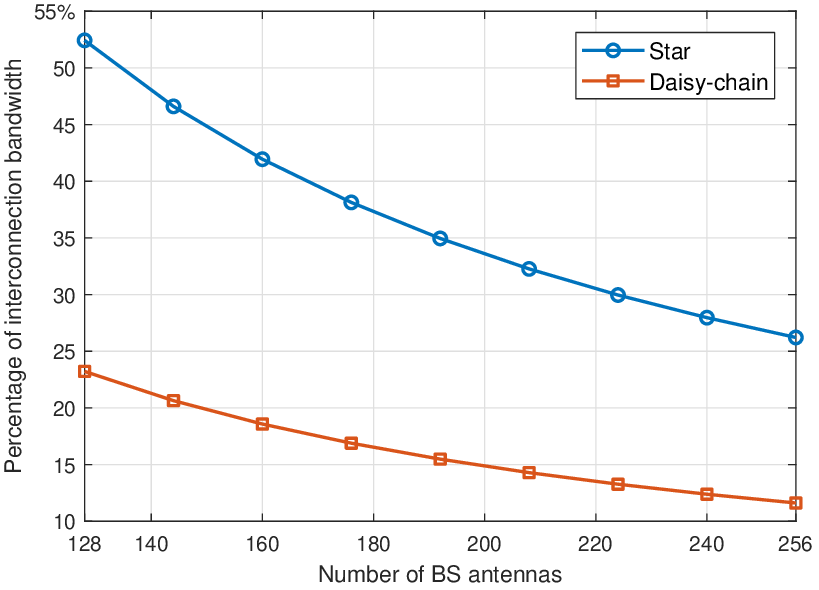}
    \caption{Percentage of interconnection bandwidth of the proposed Mini-NAG-MCMC ($m=2$ and $S=4$) relative to the centralized algorithm as a function of the number of BS antennas $B$. The MIMO system configuration is $U = 8$ and $C=8$ with 16-QAM modulation under {Rayleigh fading} channels.}
    \label{fig:interconnect2}
  \end{figure}
}

\vspace{-0.2cm}
\subsection{Interconnection Bandwidth Comparison}
We evaluate the interconnection cost of the proposed Mini-NAG-MCMC in this subsection.
The bit-width is selected as $\omega=16$ in the evaluation.
The BER performance versus interconnection bandwidth of Mini-NAG-MCMC with star and daisy-chain topologies for a MIMO system with $B=256$, $U=8$, and $C=8$ is presented in \figref{fig:interconnect1}. 
The mini-batch size for Mini-NAG-MCMC is set to $m=2$.
In the figure, the $x$-axis represents the percentage of interconnection bandwidth of Mini-NAG-MCMC relative to the centralized algorithm, 
calculated according to \eqref{eq:interconnect_centralized}-\eqref{eq:interconnect_chain} by setting the number of sampling iterations $S$ to $[2, 4, 6, 8]$.
The $y$-axis represents the performance gap between Mini-NAG-MCMC and the centralized ML algorithm in terms of the SNR required to achieve a target $\text{BER}=10^{-3}$.
From the figure, we have the following observations:
\begin{itemize}
  \item Mini-NAG-MCMC achieves a flexible trade-off between performance and interconnection bandwidth. 
  \item The daisy-chain topology provides lower interconnection bandwidth than the star topology and is more appreciated in interconnection bandwidth-constrained scenarios, despite a larger processing delay. 
\end{itemize}

Furthermore, \figref{fig:interconnect2} presents the interconnection bandwidth of  Mini-NAG-MCMC compared with the centralized algorithm  as a function of the number of BS antennas $B$ when $U=8$ and $C=8$. 
The mini-batch size and the number of sampling iterations for Mini-NAG-MCMC are $m=2$ and $S=4$, respectively.  
The figure shows that the proposed method with both star and daisy-chain topologies exhibits significantly reduced interconnection cost compared to the centralized processing as $B$ increases.
This result confirms the scalability of the proposed method to massive MIMO systems with a large number of BS antennas.

\vspace{-0.2cm}
\subsection{Trade-off between Performance and Complexity}

\CheckRmv{
  \begin{figure}[t]
    \setlength{\abovecaptionskip}{-0.2cm}
    \centering
    \includegraphics[width=3.5in]{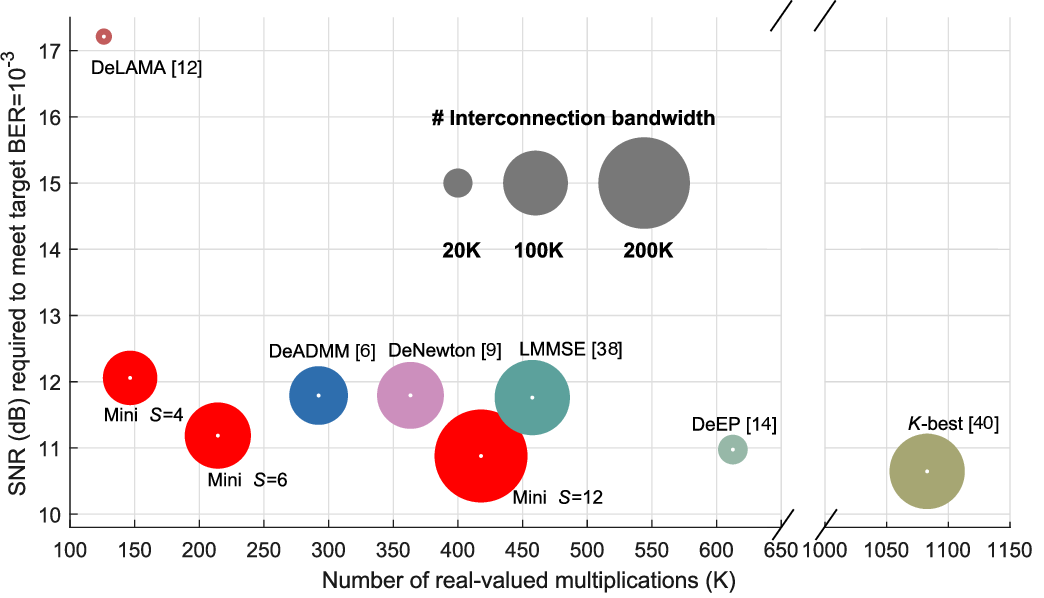}
    \caption{Performance-complexity trade-off for a MIMO system configuration of $B = 128$, $U = 32$, and $C=4$ with 16-QAM modulation under {Rayleigh fading} channels. 
    For decentralized detectors, the $x$-axis represents the cumulative count of real-valued multiplications performed at the CU along with a single DU.
    The size of the circles represents the interconnection bandwidth. 
    For the compared decentralized baselines, the interconnection bandwidth is calculated  
    by counting the entities that need to be transferred in a similar way as the analysis in \secref{sec:interconnection}.}
    \label{fig:tradeoff}
  \end{figure}
}

\figref{fig:tradeoff} presents the performance-complexity trade-off of various detectors
for a massive MIMO system with $B=128$, $U=32$, $C=4$, 16-QAM modulation, and {Rayleigh fading} channels. 
All the detectors are implemented using the star DBP topology.
The figure reveals that the proposed Mini-NAG-MCMC {($m=2$)} achieves the most promising results among the considered detectors in the trade-off between BER performance, computational complexity, and interconnection bandwidth. 
Despite the higher interconnection bandwidth of Mini-NAG-MCMC compared to DeLAMA and DeEP, 
the gap can be narrowed by deploying the proposed detector using the daisy-chain topology.
This result highlights the potential of the proposed method as a highly implementable and  desirable solution for massive MIMO detectors under DBP architectures.  

\section{Conclusion}
In this paper, the gradient-based MCMC method has been utilized for developing a novel decentralized detector for massive MIMO systems operating within the DBP architecture. 
Our approach involves decentralizing the gradient-based MCMC technique, distributing the computationally demanding gradient computations among multiple DUs for parallel processing. 
These DUs are coordinated by a CU that engages in MCMC sampling to derive a list of important samples for MIMO detection.
This design enables a balanced allocation and efficient use of computational resources. 
We have also incorporated the concept of mini-batch GD in the design of the decentralized detector, leading to reduced computation and interconnection demands across DUs while preserving high performance.  
Through extensive evaluation, we have demonstrated the clear superiority of the proposed Mini-NAG-MCMC over existing decentralized detectors, notably in terms of BER. 
Furthermore, a complexity analysis solidifies the proposed method's edge in effectively balancing performance, computational overhead, and interconnection bandwidth.

\appendix[Convergence of the Proposed Algorithm]
We illustrate the proposed algorithm's convergence in a simplified setting where the algorithm uses $\mathbf{M}_{\rm c}=\mathbf{I}$, as employed in the experiments. 
Specifically, at each sampling iteration, a candidate state $\mathbf{x}^{\prime}$ is generated as follows 
	\CheckRmv{
		\begin{equation}
			\mathbf{x}^{\prime} = Q(\mathbf{x} - \tau \nabla f_{\mathcal{I}}(\mathbf{x}) + \gamma \mathbf{w}),
			\label{eq:appendix-update}
		\end{equation}
	}
	where $\mathbf{x}$ denotes the previous state.
	By setting the global objective function as $f(\mathbf{x}) \triangleq \frac{1}{2}\|\mathbf{y}-\mathbf{H}\mathbf{x}\|^2$, we target the tempered posterior distribution given by
	\CheckRmv{
	\begin{align}
		\pi(\mathbf{x}) &\propto p(\mathbf{x})p(\mathbf{y}|\mathbf{x})^{1/T_{\rm  p}} \nonumber \\
		&\propto p(\mathbf{x})\exp\left(-{\|\mathbf{y}-\mathbf{Hx}\|^2}\right),
		\label{eq:appendix-posterior}
	\end{align}
	}	
	where $T_{\rm p}=1/\sigma^2$ is the temperature parameter \cite{farhang-boroujenyMarkovChainMonte2006,hassibiOptimizedMarkovChain2014}.
	Assuming that the prior distribution $p(\mathbf{x})$ is equiprobable, the standard MH criterion \cite{hastings1970monte} calculates the acceptance probability of $\mathbf{x}^{\prime}$ as
	\CheckRmv{
		\begin{align}
			A(\mathbf{x}^{\prime}|\mathbf{x}) &= \min \left\{1,\frac{\pi(\mathbf{x}^{\prime}) q (\mathbf{x}|\mathbf{x}^{\prime})}{\pi(\mathbf{x})q(\mathbf{x}^{\prime}|\mathbf{x})}\right\} \nonumber \\
			&= \min \left\{1, \frac{\exp (-\|\mathbf{y}-\mathbf{H} \mathbf{x}^{\prime}\|^{2})q (\mathbf{x}|\mathbf{x}^{\prime})}{\exp (-\|\mathbf{y}-\mathbf{H} \mathbf{x}\|^{2})q(\mathbf{x}^{\prime}|\mathbf{x})}\right\}.
			\label{eq:appendix-acc}
		\end{align}
	}
	However, the proposal probability $q(\cdot|\cdot)$ in \eqref{eq:appendix-acc}, which corresponds to \eqref{eq:appendix-update}, is computationally expensive. Specifically, considering the discrete space $\mathcal{A}^{U\times1}$, the proposal probability $q(\mathbf{x}^{\prime}|\mathbf{x})$ can be expressed as \cite{zhangLangevinlikeSamplerDiscrete2022}
	\CheckRmv{
		\begin{align}
		q&(\mathbf{x}^{\prime} | \mathbf{x}) = \frac{\exp\left(-\frac{1}{\gamma^2}\|\mathbf{x}^{\prime}-\mathbf{x} + \tau \nabla f_{\mathcal{I}}(\mathbf{x}) \|^2\right)}{Z_{\mathcal{A}}(\mathbf{x})}\prod_{u=1}^U \mathbb{I}_{x_u^{\prime} \in \mathcal{A}},
		\label{eq:appendix-prop}
		\end{align} 
	} 
	where $x_u^{\prime}$ is the $u$-th element of $\mathbf{x}^{\prime}$, $\mathbb{I}_{x_u^{\prime} \in \mathcal{A}}$ denotes an indicator function that returns one only if $x_u^{\prime} \in \mathcal{A}$ and zero otherwise, and $Z_{\mathcal{A}}(\mathbf{x})$ denotes a normalization constant given by
	\CheckRmv{
		\begin{equation}
			Z_{\mathcal{A}}(\mathbf{x}) = \sum_{\mathbf{x}^{\prime} \in \mathcal{A}^{U\times 1}} \exp \Big(-\frac{1}{\gamma^2}\|\mathbf{x}^{\prime}-\mathbf{x} + \tau \nabla f_{\mathcal{I}}(\mathbf{x}) \|^2\Big),
			\label{eq:appendix-Z}
		\end{equation}
	}
	which is generally intractable due to the requirement for traversal over the entire space of size $M^U$. 

	To address this challenge and simplify computation, we choose to omit the proposal probability ratio $\frac{q (\mathbf{x}|\mathbf{x}^{\prime})}{q(\mathbf{x}^{\prime}|\mathbf{x})}$ in \eqref{eq:appendix-acc}, by noting that $\nabla f_{\mathcal{I}}(\mathbf{x}) \to \mathbf{0}$ at the latter iterations of the algorithm as the chain reaches the high probability region of the state space. Given mild assumptions from \cite{wuMinibatchMetropolisHastingsMCMC2019}, this strategy is supported by the following proposition. 
	\begin{proposition}
    Assume the following:
    \begin{enumerate}
      \item $\nabla f_{\mathcal{I}}(\mathbf{x})= \mathbf{0}$.
      
      \item The learning rate $\tau = o(1)$ for large $U$, where $o(\cdot)$ denotes the standard asymptotic notation for higher-order infinitesimals \cite[Ch.~3.1]{cormen2001introduction}.
      
      \item There exists a $\lambda > 0$ such that $\sup_{(\mathbf{x},\mathcal{I})} \left\| H_{\mathcal{I}}(\mathbf{x}) \boldsymbol{z} \right\| \leq \lambda\|\boldsymbol{z}\|$ for all $\boldsymbol{z}\in \mathbb{C}^{U\times 1}$, 
      where:
      \begin{itemize}
          \item $\sup_{(\mathbf{x},\mathcal{I})}$ denotes the supremum over all possible selections of $(\mathbf{x},\mathcal{I})$,
          \item $H_{\mathcal{I}}(\mathbf{x})$ denotes the Hessian matrix on mini-batch $\mathcal{I}$.
      \end{itemize}
    \end{enumerate}
    
    Then, we have
    \CheckRmv{
      \begin{equation}
        \frac{q (\mathbf{x}|\mathbf{x}^{\prime})}{q(\mathbf{x}^{\prime}|\mathbf{x})}=1+o(1)\approx 1.
      \end{equation}
    }
	\end{proposition}
	\begin{IEEEproof}
		Defining $\mathbf{x}^{\prime}-\mathbf{x}\triangleq\boldsymbol{z}$ and by the assumption $\nabla f_{\mathcal{I}}(\mathbf{x})= \mathbf{0}$, we have
		\CheckRmv{
			\begin{align}
				\frac{q (\mathbf{x}|\mathbf{x}^{\prime})}{q(\mathbf{x}^{\prime}|\mathbf{x})} &= \frac{\exp\left(-\frac{1}{\gamma^2}\|\mathbf{x}-\mathbf{x}^{\prime} + \tau \nabla f_{\mathcal{I}}(\mathbf{x}^{\prime}) \|^2\right)Z_{\mathcal{A}}(\mathbf{x})}{\exp\left(-\frac{1}{\gamma^2}\|\mathbf{x}^{\prime}-\mathbf{x} + \tau \nabla f_{\mathcal{I}}(\mathbf{x}) \|^2\right) Z_{\mathcal{A}}(\mathbf{x}^{\prime})}  \nonumber \\
				&= \frac{\exp\left(-\frac{1}{\gamma^2}\|\tau \nabla f_{\mathcal{I}}(\mathbf{x}^{\prime}) - \boldsymbol{z} \|^2\right)Z_{\mathcal{A}}(\mathbf{x})}{\exp\left(-\frac{1}{\gamma^2}\|\boldsymbol{z} \|^2\right) Z_{\mathcal{A}}(\mathbf{x}^{\prime})}.
				\label{eq:appendix-prop-ratio}
			\end{align}
		}
		Note that 
		\CheckRmv{
			\begin{align}
				&\|\tau \nabla f_{\mathcal{I}}(\mathbf{x}^{\prime}) - \boldsymbol{z} \|^2  \nonumber \\
				= &{\tau^2} \|\nabla f_{\mathcal{I}}(\mathbf{x}^{\prime})\|^2 +\|\boldsymbol{z}\|^2 - {\tau}\big(\nabla f_{\mathcal{I}}(\mathbf{x}^{\prime})^{\rm H}\boldsymbol{z} + \boldsymbol{z}^{\rm H} \nabla f_{\mathcal{I}}(\mathbf{x}^{\prime})\big).
				\label{eq:appendix-norm}
			\end{align}
		}
		Using the Taylor series expansion $\nabla f_{\mathcal{I}}(\mathbf{x}^{\prime})= \nabla f_{\mathcal{I}}(\mathbf{x}) + H_{\mathcal{I}}(\mathbf{x}_0)(\mathbf{x}^{\prime} - \mathbf{x})= H_{\mathcal{I}}(\mathbf{x}_0) \boldsymbol{z}$ for some $\mathbf{x}_0$ between $\mathbf{x}$ and $\mathbf{x}^{\prime}$, and by the assumption $\sup_{(\mathbf{x},\mathcal{I})} \left\| H_{\mathcal{I}}(\mathbf{x}) \boldsymbol{z} \right\| \leq \lambda\|\boldsymbol{z}\|$, 
		we have 
		\CheckRmv{
			\begin{align}
				\|\nabla f_{\mathcal{I}}(\mathbf{x}^{\prime})\|^2 &= \|H_{\mathcal{I}}(\mathbf{x}_0) \boldsymbol{z}\|^2 \leq \lambda^2 \|\boldsymbol{z}\|^2, \\
				|\nabla f_{\mathcal{I}}(\mathbf{x}^{\prime})^{\rm H}\boldsymbol{z}| &= |\boldsymbol{z}^{\rm H}\nabla f_{\mathcal{I}}(\mathbf{x}^{\prime})| \leq \|\nabla f_{\mathcal{I}}(\mathbf{x}^{\prime})\| \|\boldsymbol{z}\| \nonumber\\
				&= \|H_{\mathcal{I}}(\mathbf{x}_0) \boldsymbol{z}\| \|\boldsymbol{z}\| \leq \lambda \|\boldsymbol{z}\|^2, \label{eq:appendix-ineq}
			\end{align}
		}
		where the first inequality in \eqref{eq:appendix-ineq} follows from the Cauchy-Schwarz inequality.
		Then by the assumption $\tau = o(1)$ for large $U$, which is approximately satisfied by the learning rate selection in \eqref{eq:approx_lr}, it follows that \eqref{eq:appendix-norm} is of the same order as $(1+\mathcal{O}(\tau))\|\boldsymbol{z}\|^2= (1+o(1)) \|\boldsymbol{z}\|^2$. 
		Therefore, the proposal probability ratio in \eqref{eq:appendix-prop-ratio} becomes
		\CheckRmv{
			\begin{equation}
				\frac{q (\mathbf{x}|\mathbf{x}^{\prime})}{q(\mathbf{x}^{\prime}|\mathbf{x})} = \frac{\exp\left(-\frac{1}{\gamma^2}\| \boldsymbol{z} \|^2\right)Z_{\mathcal{A}}(\mathbf{x})}{\exp\left(-\frac{1}{\gamma^2}\|\boldsymbol{z} \|^2\right) Z_{\mathcal{A}}(\mathbf{x}^{\prime})}(1+o(1))  \approx 1,
			\end{equation}
		}
    completing the proof.
	\end{IEEEproof}
	
	After omitting the proposal probability ratio, we derive the modified acceptance probability as follows
	\CheckRmv{
		\begin{equation}
			A(\mathbf{x}^{\prime}|\mathbf{x}) = \min \left\{1, \frac{\exp (-\|\mathbf{y}-\mathbf{H} \mathbf{x}^{\prime}\|^{2})}{\exp (-\|\mathbf{y}-\mathbf{H} \mathbf{x}\|^{2})}\right\},
			\label{eq:appendix-acc-approx}
		\end{equation}
	}
	which we used in \eqref{eq:acceptance2} for our proposed algorithm.
	By accepting $\mathbf{x}^{\prime}$ as the new state with this probability, the Markov chain approximately satisfies detailed balance, which is a sufficient condition to align the stationary distribution of the Markov chain with the target posterior distribution $\pi$ \cite{bishopPatternRecognitionMachine,doucetMonteCarloMethods2005}.

	Moreover, for any two states $\mathbf{x},\mathbf{x}^{\prime} \in \mathcal{A}^{U\times1}$, the transition probability of the Markov chain is given by
	\CheckRmv{
		\begin{align}
		  &T(\mathbf{x}^{\prime}|\mathbf{x}) \nonumber \\
      &=\left\{\begin{array}{ll}
			q(\mathbf{x}^{\prime} | \mathbf{x}) A(\mathbf{x}^{\prime} | \mathbf{x}), & \text { if } \mathbf{x}^{\prime} \neq \mathbf{x}, \\
			q(\mathbf{x} | \mathbf{x})+\sum_{\mathbf{u} \neq \mathbf{x}} q(\mathbf{u} | \mathbf{x})\big(1-A(\mathbf{u} | \mathbf{x})\big), & \text { otherwise},
			\end{array}\right.
		  \label{eq:appendix-trans_prob}
		\end{align}
	}
	where the second branch holds because the chain remains in state $\mathbf{x}$ either when the candidate is $\mathbf{x}$ or if the candidate is $\mathbf{u}\neq\mathbf{x}$ but gets rejected.
	With this transition probability, it is easy to show that 
	\CheckRmv{
		\begin{equation}
			T\big(\mathbf{x}_{t}=\mathbf{x}^{\prime}|\mathbf{x}_{t-1}= \mathbf{x}\big) > 0,\; \forall \mathbf{x}, \mathbf{x}^{\prime}\in \mathcal{A}^{U\times1},
		\end{equation}
	}
	implying the chain's irreducibility by definition \cite{norris1998markov}, and
	\CheckRmv{
		\begin{equation}
			\text{gcd}\left\{m: T\big(\mathbf{x}_{t+m}=\mathbf{x}|\mathbf{x}_{t}= \mathbf{x}\big) >0\right\} = 1, \; \forall m \in \mathbb{Z}^{+},
		\end{equation}
	}
	where $\mathbb{Z}^{+}$ denotes the set of positive integers, and ``gcd'' represents the greatest common divisor, verifying the chain's aperiodicity by definition \cite{norris1998markov}.
	Using these properties, it can be shown that $\pi$ is a unique stationary distribution of the proposed algorithm's underlying Markov chain \cite[Corollary 1.17]{levinMarkovChainsMixing}. This chain approximately converges to the target posterior distribution $\pi$ according to the convergence theorem of MCMC \cite[Th. 4.9]{levinMarkovChainsMixing}, thereby verifying the convergence of the proposed algorithm. 



\ifCLASSOPTIONcaptionsoff
  \newpage
\fi




\end{document}